\documentclass[aps,prd,twocolumn]{revtex4-1}
\usepackage{amsmath,amssymb,graphicx,graphics,color}

\begin{document}

\newcommand{\mcv}{$\textrm{McV}{\!}_4\,$}

\title{Particle and photon orbits in McVittie spacetimes}
\author{Brien C Nolan}
\affiliation{School of Mathematical Sciences,
Dublin City University, Glasnevin, Dublin 9, Ireland.}
\date{\today}

\begin{abstract}
McVittie spacetimes represent an embedding of the Schwarzschild field in isotropic cosmological backgrounds. Depending on the scale factor of the background, the resulting spacetime may contain black and white hole horizons, as well as other interesting boundary features. In order to further clarify the nature of these spacetimes, we address this question: do there exist bound particle and photon orbits in McVittie spacetimes? Considering first circular photon orbits, we obtain an explicit characterization of all McVittie spacetimes for which such orbits exist: there is a 2-parameter class of such spacetimes, and so the existence of a circular photon orbit is a highly specialised feature of a McVittie spacetime. However, we prove that in two large classes of McVittie spacetimes, there are bound particle and photon orbits: future-complete non-radial timelike and null geodesics along which the areal radius $r$ has a finite upper bound. These geodesics are asymptotic at large times to circular orbits of a corresponding Schwarzschild or Schwarzschild-de Sitter spacetime. The existence of these geodesics lays the foundations for and shows the theoretical possibility of the formation of accretion disks in McVittie spacetimes. We also summarize and extend some previous results on the global structure of McVittie spacetimes. The results on bound orbits are established using centre manifold and other techniques from the theory of dynamical systems.
\end{abstract}

\maketitle

\newtheorem{theorem}{Theorem}
\newtheorem{proposition}{Proposition}
\newtheorem{corollary}{Corollary}
\newtheorem{definition}{Definition}
\newtheorem{lemma}{Lemma}
\newcommand{\bi}{\begin{itemize}}
\newcommand{\ei}{\end{itemize}}
\newcommand{\be}{\begin{equation}}
\newcommand{\ee}{\end{equation}}
\newcommand{\bes}{\begin{eqnarray*}}
\newcommand{\ees}{\end{eqnarray*}}
\newcommand{\beq}{\begin{eqnarray}}
\newcommand{\eeq}{\end{eqnarray}}
\newcommand{\tao}{\tau_0}
\newcommand{\tac}{\tau_{c1}}
\newcommand{\tacc}{\tau_{c2}}
\newcommand{\real}{\mathbb{R}}
\newcommand{\rat}{\mathbb{Q}}
\newcommand{\dig}{\mathbb{D}}
\newcommand{\integer}{\mathbb{Z}}
\newcommand{\ds}[1]{\displaystyle{#1}}
\newcommand{\pd}[2]{\frac{\partial #1}{\partial #2}}
\newcommand{\poo}{\partial\Omega_0}
\newcommand{\pom}{\partial\Omega_{2m}}
\newcommand{\vx}{\vec{x}}
\newcommand{\co}{{\cal{O}}}
\newcommand{\vep}{V_\epsilon}
\newcommand{\aep}{A_\epsilon}
\newcommand{\bep}{B_\epsilon}
\newcommand{\wep}{W_\epsilon}
\newcommand{\ui}{u_{\rm{isco}}}

\newcounter{comment}
\newenvironment{comment}[1][\thedefinition]{%
      \stepcounter{comment}%
      \pagebreak[2]\medskip\par\noindent%
      \textbf{Comment \arabic{comment}}%
      \,\,\nopagebreak%
			}\smallskip\par%


%
%
%


\section{Introduction and summary.}


Perhaps one of the most striking features of black holes is their ability to create circular photon orbits: by travelling to the vicinity of a Schwarzschild black hole and settling in an orbit at a radius $r=3m$, where $m$ is the mass of the black hole, a physicist can exploit the existence of circular photon orbits at this radius to look at the back of their own head. If a Schwarzschild black hole is not available, a charged or rotating black hole can be used for the same purpose. See e.g.\ Sections 20, 40 and 61 of \cite{chandrasekhar1998mathematical}. Indeed circular photon orbits exist in Kerr-Newman-(anti)de Sitter spacetime for all parameter values corresponding to black holes \cite{stuchlik2000equatorial}. Thus it seems reasonable to say that the existence of circular photon orbits (CPOs) is a generic feature of black holes.

Similarly, in the case of massive particles, the existence of an innermost stable circular orbit (ISCO) is a feature of black holes that is not generally present in Newtonian gravity (but see \cite{amsterdamski2002marginally}). The existence of the ISCO is central to the formation of thin accrection disks around the black hole which, in turn, encodes useful information about the black hole. See \cite{lrr-2013-1}. 

The question arises as to whether or not CPOs and ISCOs arise in black holes in more general, non-vacuum settings. A first step here is to consider how such black holes can be modelled, and this question has arisen in the broader discussion of how the background expansion of the universe affects local systems \cite{faraoni2007cosmological} and in the more wide-ranging study of inhomogeneous cosmological models (see e.g.\ \cite{bolejko2011inhomogeneous}). An early contribution to this discussion was McVittie's discovery of an intriguing solution of Einstein's equations that the author himself referred to as a ``mass-particle in an expanding universe" \cite{mcvittie1933mass}. McVittie's solution represents an embedding of the Schwarzschild field in an isotropic cosmological background. In the form presented in \cite{mcvittie1933mass}, three families of line element were given, corresponding to the curvature index $k=0,\pm1$ of the isotropic background. There appear to be clear reasons to dispute the interpretation that McVittie's solutions with $k=\pm1$ represent some form of isolated system in an isotropic background \cite{nolan1998point,nandra2012effect}, and so we will focus on the spatially flat case $k=0$ \footnote{We note that both \cite{nolan1998point} and \cite{nandra2012effect} identify solutions of the Einstein equations that do for the $k=-1$ isotropic spacetimes what the McVittie spacetime does for the $k=0$ class. In comoving coordinates, these solutions are given in terms of elliptic integrals, and so are not readily amenable to the analytic studies that have been done in the $k=0$ case. As far as the author is aware, the possibility that these solutions have an elementary form in area radial coordinates has not been investigated: this may make the solutions more tractable.}. In this case, the line element may be written in the form \cite{nolan1999point}
\be ds^2=-(f-r^2H^2)dt^2-2rHf^{-1/2}dtdr+f^{-1}dr^2+r^2d\Omega^2,\label{eq:lel-mcv}\ee
where $m>0$ is a constant, $f(r)=1-2m/r>0$ (i.e.\ we restrict to $r>2m$) and $H=H(t)$ is the Hubble function of the isotropic background. That is, 
\be H(t)=S'(t)/S(t),\label{hubble}\ee where $S(t)$ is the scale factor $S(t)$ of the spacetime obtained by setting $m=0$ in (\ref{eq:lel-mcv}):
\begin{eqnarray}
ds^2 &=& -(1-r^2H^2)dt^2 -2rHdtdr+dr^2+r^2d\Omega^2 \label{eq:lel-rw1}\\
&=&-dt^2+S^2(dx^2+x^2d\Omega^2). \label{eq:lel-rw2}
\end{eqnarray}
The spacetime with this line element  will be referred to as the \textit{background}, and we will use terms such as \textit{background metric} in the obvious way.
We note that $t$ is a global time coordinate on the spacetime with line element (\ref{eq:lel-mcv}): $g^{ab}t_{,a}t_{,b}=-f^{-1}<0$. We set the time orientation of the spacetime by taking $t$ to increase into the future.
In (\ref{eq:lel-rw2}), $x=rS^{-1}$ is the usual co-moving radial coordinate of the isotropic spacetime. (We note that in (\ref{hubble}) and throughout the paper, a prime means derivative with respect to argument.) 

The limit $m=0$ of (\ref{eq:lel-mcv}) is well-defined, as it corresponds to setting the Weyl tensor of the spacetime to zero. The invariantly defined Newman-Penrose Weyl scalar $\Psi_2$ of the spherically symmetric line element (\ref{eq:lel-mcv}) is given by 
\be \Psi_2 = -\frac{m}{r^3}.\label{weyl} \ee

With $H=0$, (\ref{eq:lel-mcv}) is the line element of the Schwarzschild exterior. With $H=H_0=$ constant, the line element is that of Schwarzschild-de Sitter spacetime with cosmological constant $\Lambda=3H_0^2$. These limits are likewise well-defined, being respectively the vacuum and Einstein-space limits of (\ref{eq:lel-mcv}). In the case $H=H_0$, the coordinate transformation $T=t+u(r)$ with $u'(r)=H_0r/(\sqrt{f}(f-r^2H_0^2))$ yields the more familiar form
\be ds^2=-(1-\frac{2m}{r}-\frac{\Lambda}{3}r^2)dT^2-(1-\frac{2m}{r}-\frac{\Lambda}{3}r^2)^{-1}dr^2+r^2d\Omega^2.\ee

Imposing the Einstein equations with a cosmological constant, and with a perfect fluid energy-momentum tensor, yields expressions for the density $\rho$ and pressure $p$:
\begin{eqnarray}
8\pi\rho &=& 3H^2-\Lambda,\label{eq-rho}\\
8\pi p&=& -2H'f^{-1/2}-3H^2+\Lambda. \label{eq-press}
\end{eqnarray}
Setting $m=0$ yields the background density and pressure:
\begin{eqnarray}
8\pi\rho_0 &=& 3H^2-\Lambda,\label{eq-rho0}\\
8\pi p_0&=& -2H'-3H^2+\Lambda. \label{eq-press0}
\end{eqnarray}
Assuming that $H'(t)\neq0$, we see that McVittie spacetimes always have a scalar curvature singularity at $r=2m$.

\begin{comment}
It should be noted that since $\rho=\rho(t)$ and $p=p(t,r)$, there is no equation of state of the form $p=g(\rho)$ in the McVittie spacetime. Thus the term ``perfect fluid" is not fully appropriate: we use it in the not uncommon sense of a fluid with isotropic pressure - the radial and tangential pressures are equal. On the other hand, as the background $m=0$ spacetime is homogeneous and isotropic, one can appeal to an equation of state to close the Einstein equations and so govern the evolution. This is the perspective that we take: the insertion of the mass parameter $m$ is done \textit{post hoc}, after the Hubble function has been determined. In theory, one could then consider back reaction effects, redoing CMB and other calculations to take account of the mass parameter $m$. As we will argue below, this mass parameter provides a model of a (highly) localized inhomogeneity in an otherwise isotropic universe (we use the qualification to reflect the fact that the energy density of the universe is unaffected by $m$: $\rho=\rho(t)=\rho_0(t)$ - there is no `additional' matter in the universe).
\end{comment}

We note that the transformation $(t,r)\to (T,x)$ with $T=t$ and
\be r(T,x)= u(1+\frac{m}{2u})^2,\quad u(T,x)=xa(T) \ee
can be used to write the line element (\ref{eq:lel-mcv}) in comoving coordinates, which is the form originally given by McVittie \cite{mcvittie1933mass}. In these coordinates, the curvature singularity at $r=2m$ arises at $u=m/2$. The coordinate transformation is a diffeomorphism only if we restrict $u$ to either $(0,m/2)$ or $(m/2,\infty)$. Either interval provides a full cover of $r\in(2m,\infty)$. Note also that at fixed $T$, $\lim_{x\to \infty}r(T,x)=\infty$.

With these properties, it is tempting to conclude that the line element (\ref{eq:lel-mcv}) corresponds to a spacetime that contains a black hole embedded in an isotropic universe. However, this interpretation is too simplistic. A correct interpretation requires the thorough study of the global properties of the spacetime, based on an analysis of its geodesics. Perhaps the dominant theme that has emerged from the various studies along these lines is that different outcomes emerge depending on the background scale factor $S(t)$ and corresponding Hubble function $H(t)$.

As far as the author is aware, the first study of the global structure of McVittie spacetimes was undertaken by Sussman as part of a comprehensive study of the global properties of spherically symmetric, shear-free perfect fluid spacetimes \cite{sussman1988spherically}. In this work, a variety of possible global structures was identified. However, the interpretation of some of the results must be questioned as Sussman takes limits $x\to0$ and $x\to\infty$ within the same spacetimes. This does not seem possible without the inclusion of the singularity $r=2m$ as part of the spacetime (rather than part of its boundary). 

For expanding universe models with a big bang at a finite time in the past - that is, when the isotropic background (\ref{eq:lel-rw1}) has these features - the singularity $r=2m$ forms a past boundary of the spacetime. This was first pointed out in \cite{nolan1999point} for backgrounds satisfying a linear equation of state $p_0=\kappa\rho_0$ and with $\Lambda=0$, and the result was generalised in \cite{kaloper2010mcvittie} and \cite{lake2011more}. As we will see below, this feature of McVittie spacetimes is not universal, but does arise if the big bang condition $\lim_{t\to0^+}S(t)=0$ holds. When $\Lambda=0$, the singularity $r=2m$ also arises as a future boundary of the spacetime: future-pointing ingoing radial null geodesics run into this singularity in finite affine parameter time. However, as pointed out in \cite{kaloper2010mcvittie}, when there is a positive cosmological constant (and when some other technical conditions on $H(t)$ hold), ingoing radial null geodesics meet a horizon rather that this singularity. Subsequently, \cite{lake2011more} showed how the spacetime can extend through the horizon to a Schwarzschild-de Sitter spacetime. Together, these points indicate that with a background $\Lambda$-CDM model with $\Lambda>0$, McVittie spacetimes can indeed model black holes in expanding universes. In fact it was shown in \cite{lake2011more} that the boundary includes a black hole horizon and a white hole horizon. However, when $\Lambda=0$, it is very difficult to see how this interpretation can be given: all ingoing radial null geodesics either escape to infinity, or terminate at a scalar curvature singularity rather than reaching a horizon.

Thus, as emphasized in \cite{da2013expansion}, McVittie spacetimes can have a variety of global structures depending on the scale factor of the background. In some cases, including the $\Lambda$-CDM model, black and white hole horizons arise. Returning to the question of how local systems are affected by cosmological expansion, it is clear that these McVittie spacetimes provide an interesting testing ground for such questions. On the other hand, studying e.g.\ the existence of CPOs and ISCOs in McVittie spacetimes adds further to our understanding and correct interpretation of these spacetimes. So in this paper, we begin the study of particle and photon orbits in McVittie spacetimes by addressing these questions: do there exist CPOs in McVittie spacetimes? Do there exist bound particle and photon orbits in McVittie spacetimes? We obtain a complete answer to the first of these questions. That is, we explicitly determine all McVittie spacetimes that admit CPOs. Neither the $\Lambda$-CDM models of \cite{kaloper2010mcvittie} and \cite{lake2011more} nor the $\Lambda=0$ models of \cite{nolan1999point} mentioned above are among these. Such McVittie spacetimes do not possess this characteristic feature of black hole spacetimes. However, both models do have the following feature: both admit bound particle and photon orbits. That is, there are future-complete, non-radial null and timelike geodesics $\gamma$ in these spacetimes with the property that $r|_\gamma<+\infty$, and $\lim_{s\to\infty}r|_\gamma= r_c$ where $s$ is an affine parameter (or proper time) along the geodesic. The constant $r_c$ corresponds to the radius of a circular orbit in either a Schwarzschild or Schwarzschild-de Sitter background. This result is established for two classes of McVittie spacetimes which we define below. These classes include, respectively, the spacetimes of \cite{kaloper2010mcvittie} and \cite{lake2011more} and those of \cite{nolan1999point}. Thus these two classes of McVittie spacetimes both have this characteristic black hole property: particles and photons can be confined to a spatially compact region of spacetime by means of the spacetime geometry.

Before proceeding, we give a brief but detailed summary of the results of this paper, and note their relation to previous work. 

\subsection{Summary of Section II}
In Section II, we write down the relevant geodesic equations and state relevant dynamical systems results. We point out useful bounds relating $\dot{t}$ and $\dot{r}$ along individual geodesics - see (\ref{ineq}). 

\subsection{Summary of Section III}  In Section III, we derive and solve an equation that must hold for the Hubble function $H(t)$ of a McVittie spacetime that admits a CPO. We briefly discuss global properties of the associated spacetime: a fuller discussion is given in Appendix A. As we see in (\ref{eq:lel-mcv}), McVittie spacetimes are determined by a parameter $m$ and a function $H$. We show in Section III that there is (only) a 2-parameter family of McVittie spacetimes that admit a CPO. Thus this is a `no-go' result: CPOs are absent from nearly all McVittie spacetimes. We note that the same conclusion holds in relation to circular timelike orbits. 

\subsection{Summary of Section IV} In Section IV we define a class of McVittie spacetimes that we will refer to as \textit{expanding McVittie spacetimes with a big bang background}. These are the focus of the remainder of the paper: this class appears to us to be the class of McVittie spacetimes of most physical interest. See Definition 1. We prove that in this class, all causal geodesics (timelike and null, radial and non-radial) originate at the singular boundary $r=2m$ at finite affine distance (proper time) in the past. This generalises previous results relating to radial null geodesics \cite{nolan1999point,kaloper2010mcvittie,lake2011more}.

\subsection{Summary of Section V}
Here, we deal with the future evolution of causal geodesics. We define two subclasses of expanding McVittie spacetimes with a big bang background that are distinguished by the (future) asymptotic value of the Hubble function $H(t)$. For Class 1, $H(t)\to H_0>0$ as $t\to+\infty$ and for Class 2, $H(t)\to 0$. See Definitions 2 and 3. We derive rigorously and generalise results that have been presented previously either in a heuristic manner, or for special cases. In particular, we prove that the global structure (in the sense of a Penrose-Carter conformal diagram) obtained in \cite{lake2011more} for a special case - i.e.\ $H(t)$ specified - holds generally for Class 1 (we note that several of the results of \cite{lake2011more} were proven in general). Included under this heading are existence proofs relating to important radial null geodesics that delineate the global structure (see Propositions 4 - 7). The crucial role of a non-zero value of $H_0$ was first identified in \cite{kaloper2010mcvittie}: this has far-reaching consequences in that it changes radically the interpretation of the corresponding McVittie spacetime, as discussed above. In Section V, our intention, in part, is to put the physical insights of this paper on a sounder mathematical footing. We note also a generalisation and some corrections (Proposition 7 (b) - noted in \cite{lake2011more}) to the results of \cite{kaloper2010mcvittie}. For example, the linear equation of state used in Appendix A of that paper is not required. In addition, we generalise some previous results for the case $H_0=0$ \cite{nolan1999point}.

\subsection{Summary of Section VI} 
The main results of the paper are given here. Using various techniques from dynamical systems, we prove the existence of bound photon and particle orbits in both Class 1 and Class 2 spacetimes. That is, we prove the existence of future-complete non-radial timelike and null geodesics $\gamma$ with the property that $r|_\gamma$ is finite along the whole history of the geodesic. Taking $s$ to be the parameter (affine parameter or proper time) along the geodesic, we show that $\lim_{s\to+\infty} r(s) = r_c$, where $r_c$ is the radius of a circular orbit in the corresponding Schwarzschild-de Sitter (Class 1) or Schwarzschild (Class 2) spacetime. For photon orbits, this forces $r_c=3m$ and for particle orbits, $r_c$ corresponds to a stable circular orbit. See Propositions 10 and 12-14.  

We use units with $G=c=1$, and we use a $\square$ to indicate the end of a proof.


\section{General properties of causal geodesics}

We begin with the line element (\ref{eq:lel-mcv}) and define
\be \chi(t,r)=f-r^2H^2,\quad \alpha(t,r)=rf^{-1/2}H,\quad\lambda(r)=f^{-1}.\label{eq:metric-fns}\ee
Recall that $f(r)=1-2m/r$, and note that $\chi\lambda+\alpha^2=1$. The geodesic equations of the spacetime may be written in the following form:
\begin{eqnarray}
\ddot{r}&=&rf^{1/2}H'\dot{t}^2+(1-\frac{3m}{r})\frac{\ell^2}{r^3}+\epsilon(\frac{m}{r^2}-rH^2),\label{ng1}\\
\ddot{t}&=&-(1-\frac{3m}{r})f^{-1/2}H\dot{t}^2-\frac{2m}{r^2}f^{-1}\dot{t}\dot{r}-\epsilon f^{-1/2}H,\label{ng2}
\end{eqnarray}
where $\ell$ is the conserved angular momentum of the geodesic, $\epsilon=0$ for null geodesics and $\epsilon=-1$ for timelike geodesics. The overdot represents the derivative with respect to the parameter along the geodesic: an affine parameter for null geodesics and proper time for timelike geodesics. Throughout the remainder of this paper, we use $s$ to represent this parameter. We also have the first integral of (\ref{ng1})-(\ref{ng2}):
\be -\chi\dot{t}^2-2\alpha\dot{t}\dot{r}+f^{-1}\dot{r}^2+\frac{\ell^2}{r^2}=\epsilon.\label{ng3}\ee
It follows that $\dot{t}\neq0$ everywhere on a causal geodesic. Then by a choice of parameter orientation, we have $\dot{t}>0$ everywhere along all causal geodesics. 

We consider the geodesic equations on the region $\Omega=\{(t,r):t>0, r>2m\}$. This region has boundaries $\poo=\{(0,r):r\geq 2m\}$ and $\pom=\{(t,2m):t>0\}$. It will be convenient to consider the geodesic equations as a first order dynamical system. To this end, we define $\vec{x}=(r,t,\dot{r},\dot{t})$ and write the geodesic equations in the form
\be \frac{d{\vec{x}}}{ds}=\vec{F}(\vec{x}),\label{geo-ds}\ee
where $F_1(\vec{x})=\dot{r}=x_3$, $F_2(\vec{x})=\dot{t}=x_4$ and $F_3(\vec{x}), F_4(\vec{x})$ can be read-off the right hand sides of (\ref{ng1}) and (\ref{ng2}). We note that $\vec{F}\in C^1(E)$, where $E=\Omega\times \mathbb{R}^2$. In this context, (\ref{ng3}) is a zero-order constraint that is propagated by the equations. That is, if (\ref{ng3}) holds for $\vec{x}=\vec{x}_0$, then this equation is valid everywhere along the solution $\vx$ of (\ref{geo-ds}) with the initial condition $\vx(0)=\vx_0$.

Local existence of $C^1$ solutions of (\ref{geo-ds}) follows from standard results of dynamical systems (see e.g.\ \cite{perkodifferential}). For any $\vx_0\in E$, there is a maximal interval of existence $(s_\alpha,s_\omega)=(s_\alpha(\vx_0),s_\omega(\vx_0))$ of the initial value problem
\be \frac{d{\vec{x}}}{ds}=\vec{F}(\vec{x}),\quad \vx(0)=\vx_0.\label{geo-ivp}\ee
The intervals $(s_\alpha,0]$ and $[0,s_\omega)$ are respectively the left-maximal and right-maximal intervals of existence. The following theorem plays an important role in the next section:
\begin{theorem}{(Perko,\cite{perkodifferential})}\label{thm1}
Let $E$ be an open subset of $\mathbb{R}^n$ containing $\vx_0$, let $\vec{F}\in C^1(E)$ and let $(s_\alpha,0]$ be the left-maximal interval of existence of the IVP (\ref{geo-ivp}). If $s_\alpha>-\infty$, then given any compact set $K\subset E$, there exists $s\in(s_\alpha,0)$ such that $\vx(s)\not\in K$.
\end{theorem}

Roughly speaking, this theorem tells us that solutions of (\ref{geo-ivp}) continue to exist while $\vx$ remains bounded. We will exploit this theorem to prove extension results for causal geodesics.

By completing a square, we can write (\ref{ng3}) in the form
\be  \dot{t}^2-(\alpha\dot{t}-\lambda\dot{r})^2=\lambda(\ell^2r^{-2}+|\epsilon|), \ee
so that for non-radial geodesics ($\ell\neq 0$),
\be  ((1-\alpha)\dot{t}+\lambda\dot{r})((1+\alpha)\dot{t}-\lambda\dot{r})>0. \ee
Noting that $\lambda$ and $\dot{t}$ are both positive, and that $\alpha>0$ in an expanding ($H>0$) McVittie spacetime, we obtain the following:

\begin{lemma}\label{lem1}
At every point on a non-radial causal geodesic in an expanding McVittie spacetime,
\be (\alpha-1)\dot{t}<\lambda\dot{r}<(\alpha+1)\dot{t}.\label{ineq}\ee
\hfill$\square$
\end{lemma}


\section{Circular photon orbits}

For circular photon orbits, we have $r=a=$ constant along the null geodesic. We take $a>2m$. Imposing this condition in (\ref{ng1}) and (\ref{ng3}), eliminating $\ell$ and noting that $\dot{t}\neq0$, we obtain the following ODE:
\be H'(t) = A(H^2-H_0^2),\label{eq:main}\ee
where
\be A= \frac{\left(1-\frac{3m}{a}\right)}{\sqrt{1-\frac{2m}{a}}},\quad H_0=a^{-1}\sqrt{1-\frac{2m}{a}}.\label{eq:ah0def}
\ee
We note that if the circular photon orbit has radius $a=3m$, then $A=0$ and we are in Schwarzschild-de Sitter spacetime. So for a non-trivial solution, we exclude $a=3m$. We note then that $mH_0<\frac{1}{3\sqrt{3}}$ (equality holds for $a=3m$).
The only remaining condition of (\ref{ng1})-(\ref{ng3}) is
\be \chi(t,a)\dot{t}^2=\frac{\ell^2}{a^2}.\label{ng4}\ee
This yields the following result.
\begin{proposition}\label{prop:CPO1}
The McVittie spacetime with line element (\ref{eq:lel-mcv}) admits circular photon orbits if and only if the Hubble function $H(t)$ satisfies (\ref{eq:main}), with the constants $A, H_0$ determined by the mass parameter $m$ and the orbital radius $a$ via (\ref{eq:ah0def}), and $\chi(t,a)>0$ for all $t$ along the orbit. \hfill$\square$
\end{proposition}

The function $\chi$, which is invariantly defined by $\chi=g^{ab}\nabla_ar\nabla_br$, plays an important role in spherical symmetry: this leads to an interesting corollary to Proposition \ref{prop:CPO1}. As is well known, $\chi$ is proportional to the product of the expansions of the ingoing and outgoing radial null geodesics of a spherically symmetric spacetime. Hence the surface (or surfaces) $\chi=0$ corresponds to a horizon, where (at least) one of the null expansions vanishes. The \textit{regular} (or untrapped) region of the spacetime corresponds to $\chi>0$, where one null expansion is positive and one is negative, while $\chi<0$ corresponds to either a \textit{trapped} region (two negative null expansions) or an \textit{anti-trapped} region (two positive null expansions). We immediately have the following, which is a particular case of a more general result:
\begin{corollary}\label{corr:corr1}
A circular photon orbit of a McVittie spacetime is confined to the regular region of the spacetime.
\hfill$\square$
\end{corollary}

We can also see from (\ref{ng3}) that any turning points (periapsis or apasis, whereat $\dot{r}=0$) of a photon orbit in a McVittie spacetime must lie in the regular region of the spacetime. The same conclusion holds for particle orbits (timelike geodesics), where (\ref{ng3}) holds but with $-1$ on the right hand side. In fact it is readily seen that Corollary 1 holds much more generally in spherically symmetric spacetimes (this is the more general result referred to above). Using double null coordinates $(u,v)$, both taken to increase into the future, the null expansion $\theta_u$ along the future pointing null direction $\frac{\partial}{\partial u}$ satisfies $\theta_u=\frac{1}{\kappa_u^2}\frac{\partial r}{\partial u}$ for some positive function $\kappa_u^2$. A corresponding statement holds with $u$ replaced by $v$. Then along any future pointing null geodesic,
\be  \dot{r} = \kappa_u^2\theta_u\dot{u}+\kappa_v^2\theta_v\dot{v}. \ee
With $u,v$ increasing into the future, so that $\dot{u}> 0$ and $\dot{v}>0$, we see that $\dot{r}$ can vanish only when either the null expansions $\theta_u, \theta_v$ have opposite sign - that is, when the geodesic is in a regular region of the spacetime - or when both $\theta_u$ and $\theta_v$ vanish. This latter situation is non-generic: it occurs for example at the bifurcation 2-sphere of the extended Schwarzschild (Kruskal-Szekeres) spacetime. The former conclusion should hold in general.

Corollary 1 is almost enough to rule out CPO's in any McVittie spacetime for which the isotropic background has a big bang - i.e.\ $S(t_0)=0$ at some time $t_0$ in the past. From (\ref{hubble}), we see that $H(t)$ will typically diverge as $t\to t_0^+$. But then $\chi(t,a)=1-\frac{2m}{a}-a^2H^2$ becomes negative at early times. It is very difficult to see how the CPO can exit the trapped region, or terminate in the past at a finite positive value of $t$, without there being some serious pathology of the spacetime. Thus the existence of a CPO is a very strong restriction on McVittie spacetimes with big bang backgrounds.

We turn now to determining which line elements of the form (\ref{eq:lel-mcv}) are admitted by Proposition \ref{prop:CPO1}. This amounts to solving (\ref{eq:main}), and ensuring that the resulting $\chi(t,a)$ is positive.
There is a unique non-trivial solution (where non-trivial means that $H$ is not constant: recall that this corresponds to Schwarzschild-de Sitter spacetime):
\be H(t) = -H_0\tanh(AH_0t),\quad t\in \mathbb{R}.\label{eq:hsol}\ee
We have used time translation freedom to set $t=0$ at the zero of $H$. The corresponding scale factor is
\be S(t) = (\cosh(AH_0t))^{-1/A},\label{asol}\ee
and we find
\be \chi(t,a) = a^2H_0^2/\cosh^2(AH_0t).\label{chisol}\ee
There are two inequivalent spacetimes, depending on the sign of $A$. We see from (\ref{eq:ah0def}) that $A$ is positive (respectively negative) if the CPO radius $a$ is greater than (respectively less than) $3m$. The expansion histories of the background universes corresponding to the two choices are shown in Figures 1 and 2, and their global structure is analysed in the Appendix A.  

\begin{figure}\label{hubble-scale-a-neg}
		{\includegraphics[scale=0.5]{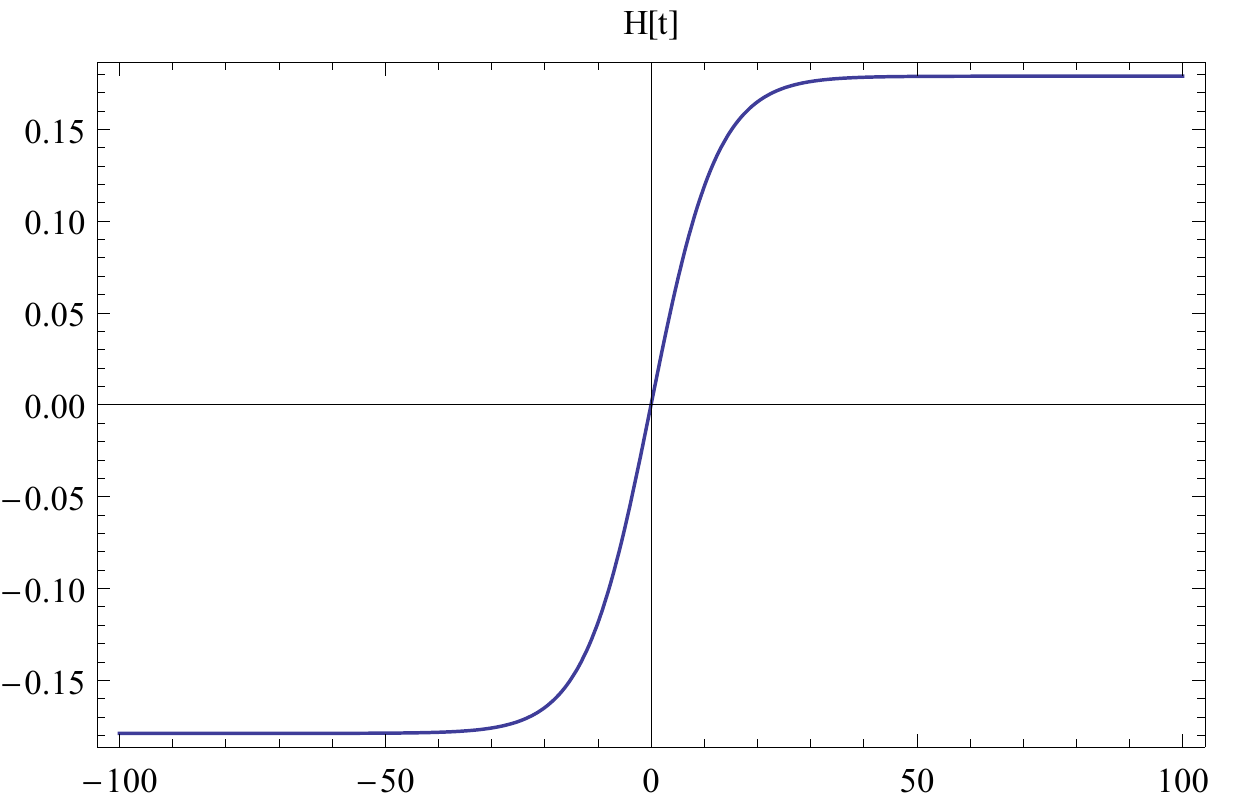} \includegraphics[scale=0.5]{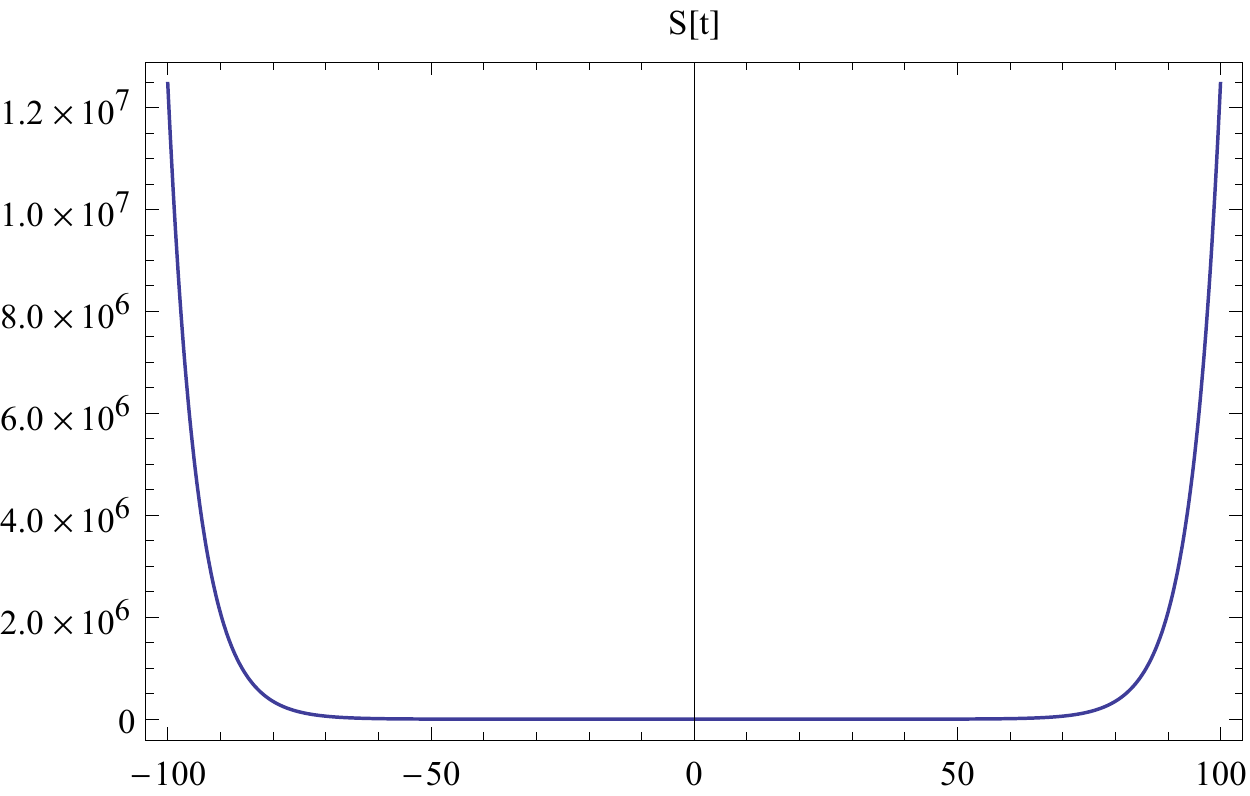}}
	\label{fig:CPO-hubble-A-negative}
	\caption{The graph of (i) the Hubble function $H(t)$ and (ii) the scale factor $S(t)$ for the McVittie spacetime admitting a circular photon orbit at radius $a<3m$. Here $m=1$ and $a=2.5$, giving $A=-0.447214<0$ and $H_0=0.178885$.}
\end{figure}
\begin{figure}\label{hubble-scale-a-pos}
		{\includegraphics[scale=0.5]{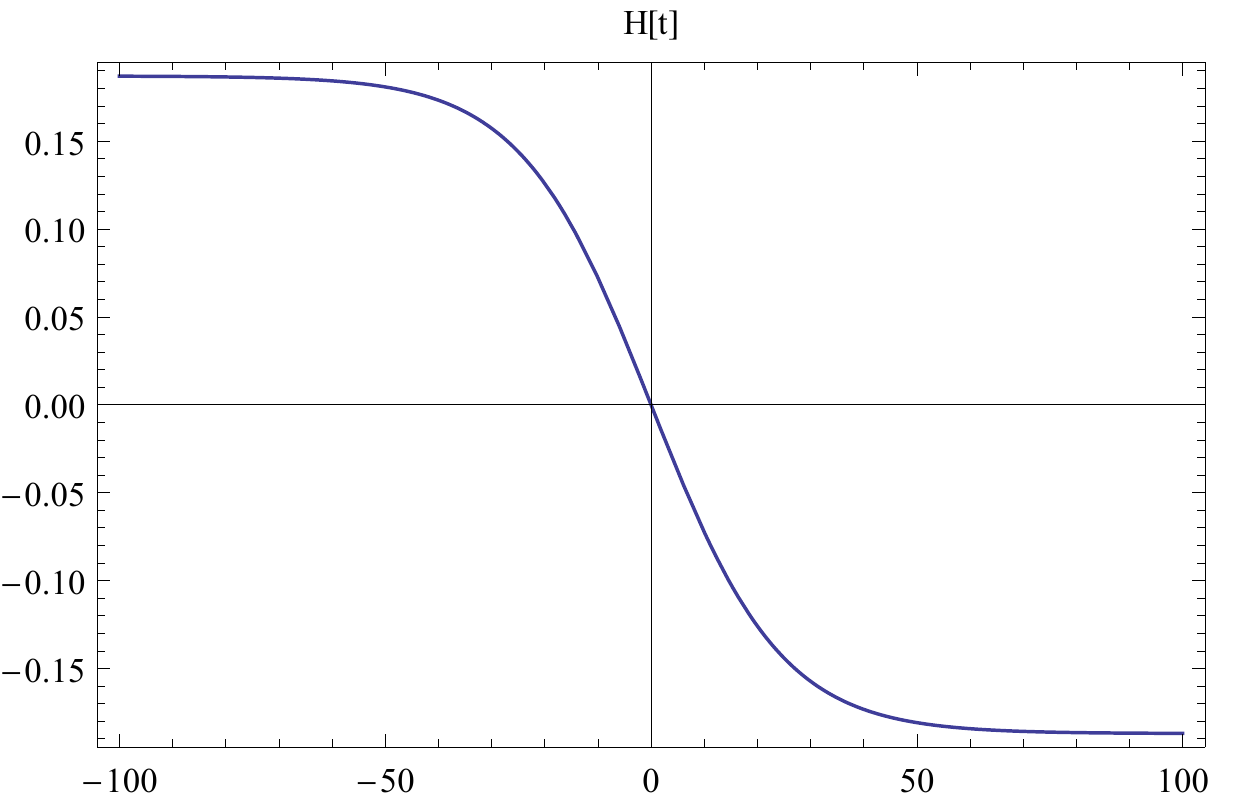} \includegraphics[scale=0.5]{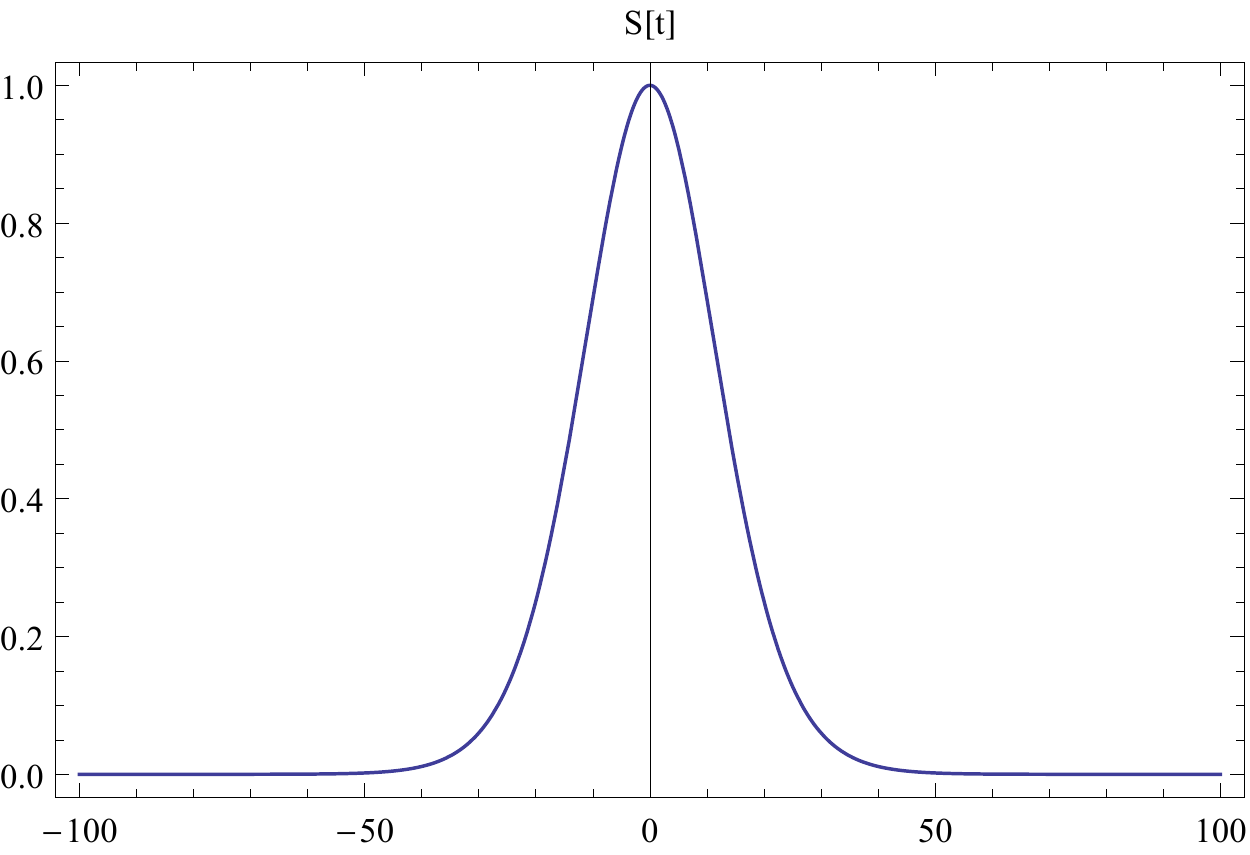}}
	\label{fig:CPO-hubble-A-positive}
	\caption{The graph of (i) the Hubble function $H(t)$ and (ii) the scale factor $S(t)$ for the McVittie spacetime admitting a circular photon orbit at radius $a>3m$. Here $m=1$ and $a=3.5$,giving $A=0.218218>0$ and $H_0=0.187044$.}
\end{figure}

We find it curious that (\ref{eq:main}) has a solution which {\it does} satisfy the $\Lambda$-CDM conditions of \cite{lake2011more} ({see also Definition 1 below}). This solution is $H(t)=-H_0\coth(AH_0t)$. However this yields $\chi(t,a)=-a^2H_0^2/\sinh^2(AH_0t)$, and so the second condition (\ref{ng4}) for a CPO is violated.

We summarise the main result of this section as follows.
\begin{proposition}\label{prop:CPO2}
The McVittie spacetime with line element (\ref{eq:lel-mcv}) admits a circular photon orbit if and only if the Hubble function is given by (\ref{eq:hsol}), with $A,H_0$ determined by the mass parameter $m$ and photon orbital radius $a$ as in (\ref{eq:ah0def}).\hfill$\square$
\end{proposition}

\begin{comment}
It is straightforward to derive and solve the equation corresponding to (\ref{eq:main}) in the case of circular \textit{particle} orbits. This yields a different McVittie spacetime to that of Proposition \ref{prop:CPO2} - but again, only a 2-parameter family of spacetimes arises. We will not pursue this further: the main point is that circular orbits almost never exist in McVittie spacetimes. 
\end{comment}


\section{Past evolution}

We begin the dicsussion by considering the past evolution of causal geodesics in the class of McVittie spacetimes of interest. These correspond to McVittie spacetimes for which the isotropic background is an expanding cosmological model - an expanding isotropic spacetime with a big bang at a finite time in the past. 

\begin{definition} The spacetime with line element (\ref{eq:lel-mcv}) is said to be an \textbf{expanding McVittie spacetime with a big bang background} if the following conditions on the Hubble function hold:
\begin{itemize}
\item[(i)] $H\in C^2((0,+\infty),\mathbb{R}_+)$;
\item[(ii)] $\lim_{t\to0^+}\int_t^{t_0} H(u)du=+\infty$ for all $t_0>0$;
\item[(iii)] $H'(t)<0$ for all $t>0$.
\end{itemize}
\end{definition}

The first condition here includes a technical differentiability condition and the condition for expansion: $H(t)>0$ for all $t>0$. The second incorporates the big bang condition. This is equivalent to the existence of some $t_0\in\mathbb{R}$ - which we set to zero by a translation - for which the scale factor $S$ satisfies $S(t_0)=0$. The third condition is equivalent to the weak energy condition $\rho+p>0$ in both the background and the McVittie spacetime - see (\ref{eq-rho}) and (\ref{eq-press}).

In this section, we establish the fact that $\pom=\{(t,r):t>0,r=2m\}$ - which, recall, is a curvature singularity - forms the past boundary of the spacetime. That is, \textit{all} causal geodesics originate in the past at $\pom$ at finite affine distance (null geodesics) or finite proper time (timelike geodesics). All such geodesics extend back to $\pom$ at a positive value of $t$: the background big bang surface $\poo=\{(t,r):t=0, r\geq 2m\}$ is cut off by $\pom$. These statements are non-trivial and so proofs are given below. We note that it is straightforward to see that $r=2m$ forms a spacelike \textit{portion} of the past boundary of the spacetime \cite{kaloper2010mcvittie}: what is not obvious is that \textit{all} causal geodesics originate here. This is stated formally as Proposition \ref{prop:causal-past} below.

The radial null geodesics of (\ref{eq:lel-mcv}) are of particular importance in determining the global structure of the spacetime. These are the geodesics satisfying (\ref{ng1})-(\ref{ng3}) with $\ell=\epsilon=0$. Noting that $t$ is a global time coordinate of (\ref{eq:lel-mcv}) that increases into the future by a choice made in Section I above, we have $\dot{t}>0$ everywhere along a causal geodesic, and so from (\ref{ng3}) we can write down
\be \frac{dr}{dt}=Hr(1-\frac{2m}{r})^{1/2}\pm(1-\frac{2m}{r})\label{rng}\ee
for radial null geodesics (RNG). There are two families: outgoing, corresponding to the upper sign, and ingoing, corresponding to the lower. We introduce the radial coordinate $z$ as defined in \cite{nolan1999point}:
\be z=\sqrt{1-\frac{2m}{r}}.\label{eq:zdef}\ee 
Notice then that $z\in[0,1)$ with $z=0\Leftrightarrow r=2m$ and $z\to1$ as $r\to\infty$. In terms of $z$, $\Omega=\{(t,z):t\geq 0, z\in(0,1)\}$ and the RNGs satisfy
\be \frac{dz}{dt}=\frac{H}{2}(1-z^2)\pm\frac{z}{4m}(1-z^2)^2.\label{rngz}\ee

Our first step is to establish the following: given any point $P:(t,z)=(t_1,z_1)\in\Omega$ (so that $t_1>0$ and $z_1\in(0,1)$), the ingoing and outgoing radial null geodesics (IRNGs and ORNGs) through $P$ both originate on the surface $\pom=\{r=2m,0<t<\infty\}=\{z=0,0<t<\infty\}$. We then show that the same holds for all causal geodesics passing through $P$, and thereby show that the set $\poo$ does not form part of the past boundary of the spacetime.

\begin{lemma}\label{lem2} In an expanding McVittie spacetime with a big bang background, all outgoing radial null geodesics of the spacetime originate at $\pom$ at finite affine distance in the past. That is, for each ORNG $\gamma$, there exists $s_0>-\infty$ such that $\lim_{s\to s_0^+}(t(s),r(s))|_{\gamma}=(t_0,2m)$ for some $t_0>0$.
\end{lemma}

\noindent\textbf{Proof:} Consider the ORNG through $(t,z)=(t_1,z_1)$ with $t_1>0$ and $z_1\in(0,1)$. This satisfies
\be  \frac{dz}{dt}=\frac{H}{2}(1-z^2)+\frac{z}{4m}(1-z^2)^2>0 \ee
for $t>0,0\leq z<1$. So for $t<t_1$, we have $z\leq z_1$, and hence
\be  \frac{dz}{dt}\geq\frac{H}{2}(1-z_1^2)+\lambda_1 z, \ee
where $\lambda_1 = (1-z_1^2)^2/4m>0$, and we recall from part (i) of Definition 1 that $H>0$. Integrating this linear differential inequality from $t$ to $t_1$ yields \begin{eqnarray*} z(t) &\leq& z_1e^{\lambda_1(t-t_1)}-\frac{(1-z_1)^2}{2}\int_t^{t_1}H(u)e^{\lambda_1(t-u)}du\\
&\leq & z_1e^{\lambda_1(t-t_1)}-\frac{(1-z_1)^2}{2}e^{-\lambda_1t_1}\int_t^{t_1}H(u)du.\end{eqnarray*}
These inequalities hold for $t<t_1$. From part (ii) of Definition 1, we see that $z(t)$ must reach 0 (and so $r$ reaches $2m$) at some positive value $t_0$ of $t$. From (\ref{ng2}) with $\ell=0$ and part (iii) of Definition 1, we see that $\ddot{r}<0$ along the geodesic. This guarantees that $r$ reaches $2m$ at a finite value $s_0$ of the affine parameter $s$. \hfill$\square$

\begin{lemma}\label{irng-lemma} In an expanding McVittie spacetime with a big bang background, all ingoing radial null geodesics of the spacetime originate at $\pom$ at finite affine distance in the past. That is, for each IRNG $\gamma$, there exists $s_0>-\infty$ such that $\lim_{s\to s_0^+}(t(s),r(s))|_{\gamma}=(t_0,2m)$ for some $t_0>0$.
\end{lemma}

\noindent\textbf{Proof:} For IRNGs, we have
\be  \frac{dz}{dt}=\frac{H}{2}(1-z^2)-\frac{z}{4m}(1-z^2)^2. \ee
We note that $0\leq z(1-z^2)\leq 2/(3\sqrt{3})$ for all $z\in[0,1)$. Then defining $H_c=\frac{1}{3\sqrt{3}m}$, we find
\be  \frac{dz}{dt}\geq \frac12(H-H_c)(1-z^2),\qquad t>0,\quad 0\leq z<1. \ee
Integrating from $(t,z)$ to $(t_1,z_1)$ where $t_1>t$ eventually yields
\be  z(t) \leq \frac{(1+z_1)-(1-z_1)e^{I(t)}}{(1+z_1)+(1-z_1)e^{I(t)}},\qquad t<t_1 \ee
where
\be  I(t) = \int_t^{t_1}(H(u)-H_c)du. \ee
Divergence of $I$ in the limit $t\to 0^+$ (which comes from part (ii) of Definition 1) shows that $z$ reaches 0 at a positive value of $t$. The proof that this occurs at a finite value of the affine parameter is identical to the corresponding proof in Lemma \ref{lem2}. \hfill$\square$

The following results establish that the results of these two lemmas apply to \textit{all} causal geodesics. We first establish a somewhat obvious `radial confinement' result.

\begin{lemma}\label{lem4} Let $p\in\Omega$ and let $\gamma$ be a causal geodesic with $\gamma(0)=p$. Then the past branch of $\gamma$ - that is, the set of points
\be \gamma^-=\{\gamma(s):s_\alpha<s<0\} \label{gam-minus}\ee
is contained in the interior of the region of $\Omega$ bounded by the point $p$, by the paths of the past-directed radial null geodesics through $p$ and by the boundary $\pom$.
\end{lemma}

\noindent\textbf{Proof:} Noting that $\dot{t}>0$ along $\gamma$, we find from (\ref{ineq}) that
\be \lambda^{-1}(\alpha-1)<\frac{dr}{dt}<\lambda^{-1}(\alpha+1).\ee
The lower and upper bounds correspond to the uniquely defined value at each point of $\Omega$ of the slopes of the ingoing and outgoing radial null geodesics. Thus at any given point of $\Omega$, the geodesic $\gamma$ crosses the ingoing (respectively outgoing) radial null geodesic from below (respectively above). It follows that to the past of $p$ (i.e.\ for $s<0$), $\gamma$ remains below (respectively above) the ingoing (respectively outgoing) radial null geodesic through $p$. The conclusion follows. \hfill$\square$

\begin{comment} It is tempting to conclude on the basis of this lemma that any causal $\gamma$ must extend back to $r=2m$. However it remains to prove that $\gamma$ extends sufficiently far into the past in order that this happens. We now prove this extension result.
\end{comment}
\begin{proposition}\label{prop:causal-past} Let $\gamma$ be a future-pointing causal geodesic of an expanding McVittie spacetime with a big bang background. Then there exists $t_\alpha>0$ and $s_\alpha >-\infty$ such that $\lim_{s\to s_\alpha^+} (t(s),r(s))|_\gamma = (t_\alpha,2m)$ where $s$ is an affine parameter (respectively proper time) along a null (respectively timelike) geodesic. That is, all causal geodesics of an expanding McVittie spacetime with a big bang background originate at $\pom$.
\end{proposition}

\noindent\textbf{Proof:} Let $\gamma(0)=p\in\Omega$, and as before, let $(s_\alpha,0]$ be the left maximal interval of existence for $\gamma$. Applying the radial confinement result of the previous lemma and recalling that $\dot{t}>0$ along $\gamma$, we have
\be t_i(p) < t(s) < t(0),\quad s\in (s_\alpha,0),\label{t-bounds}\ee
where $t_i(p)$ is the value of $t$ at which the ingoing radial null geodesic through $p$ meets $\pom$. Likewise,
\be r(s) < r_+(p),\quad s\in(s_\alpha,0),\label{r-bound}\ee
where $r_+(p)$ is the maximum of the value of $r$ along the past branches of the radial null geodesics through $p$.

From Lemma \ref{lem1}, and recalling that $\lambda=f^{-1}$ and that $\dot{t}>0$, we have
\be -\frac{2m}{r}f^{-1}\dot{t}\dot{r}>-\frac{2m}{r}(\alpha+1)\dot{t}^2.\ee
Then from (\ref{ng2}), we can write
\be \ddot{t}>-\left((1-\frac{m}{r})f^{-1/2}H+\frac{2m}{r^2}\right)\dot{t}^2,\label{tdd-lower-bound}\ee
where we note that the coefficient of $\dot{t}^2$ on the right hand side is strictly negative on $\Omega$.

For $p\in\Omega$ and $\epsilon>0$, define the compact sets
\be K_{p,\epsilon}=\{(t,r): t_i(p)-\epsilon \leq t \leq t(0), 2m+\epsilon \leq r \leq r_+(p)\}.\label{kpep}\ee
Then there exists a positive constant $C_{p,\epsilon}$ such that
\be (1-\frac{m}{r})f^{-1/2}H+\frac{2m}{r^2} \leq C_{p,\epsilon},\quad \hbox{for all } (t,r)\in K_{p,\epsilon}. \label{bound}\ee
Now suppose that the past branch of $\gamma$ is confined to $K_{p,\epsilon}$ - that is, that $(t(s),r(s))\in K_{p,\epsilon}$ for all $s\in(s_\alpha,0)$. Then along $\gamma$, 
\be \ddot{t}\geq - C_{p,\epsilon}\dot{t}^2,\quad s\in(s_\alpha,0).\label{tdd-bound}\ee
Integrating yields
\be \dot{t}(s) <\dot{t}(0)\exp[(t(0)-t(s))C_{p,\epsilon}],\quad s\in(s_\alpha,0),\label{td-bound1}\ee
and so
\be 0 \leq \dot{t}(s) \leq \dot{t}(0)\exp[t(0)C_{p,\epsilon}],\quad s\in(s_\alpha,0).\label{td-bound2}\ee
Using (\ref{ineq}), this yields finite upper and lower bounds for $\dot{r}$:
\be C_{p,\epsilon}^- \leq \dot{r}(s) \leq C_{p,\epsilon}^+,\quad s\in(s_\alpha,0),\label{rd-bound}\ee
where  the bounding constants $C_{p,\epsilon}^\pm$ depend on the parameter $\epsilon$, the initial point $p$ and the initial value and derivative of $t$.

 From (\ref{td-bound2}) and (\ref{rd-bound}), it follows that the corresponding solution $\vec{x}(s)$ of the dynamical system (\ref{geo-ds}) is confined to the compact subset $K$ of $E=\Omega\times\mathbb{R}^2$, where
\be K = K_{p,\epsilon} \times [C_{p,\epsilon}^-,C_{p,\epsilon}^+]\times[0,\dot{t}(0)\exp[t(0)C_{p,\epsilon}].\label{kdef}\ee
This contradicts Theorem \ref{thm1}, and so $\gamma$ must exit $K_{p,\epsilon}$. The radial confinement result indicates that $\gamma$ must exit the set via the lower boundary $r=2m+\epsilon$. That is, for every $\epsilon>0$, there exists $s_0\in(s_\alpha,0)$ so that $r(s_0)<2m+\epsilon$. It follows that there exists $s_{2m}\in[-\infty,0)$ such that $\lim_{s\to s_{2m}^+} r(s) = 2m$. We now show that $s_{2m}>-\infty$, completing the proof. (Note that $s_{2m}=-\infty$ corresponds to $s_\alpha=-\infty$.)

Recall that $f=1-2m/r$ and that $H(t)>0$ with $H'(t)<0$. It follows that $H(t(s))\leq H(t_i(p))$ for all $s\in(s_\alpha,0)$. Recall that we may write (\ref{ng3}) in the form
\be \dot{t}^2 - (\alpha\dot{t}-\lambda\dot{r})^2 = f^{-1}(|\epsilon|+\frac{\ell^2}{r^2}).\ee
It follows that
\be \lim_{s\to s_{2m}^+} \dot{t}(s) = +\infty.\ee
Using the left-hand inequality of (\ref{ineq}) in (\ref{ng2}), we can write
\be \ddot{t} < - (1-\frac{m}{r})f^{-1/2}H\dot{t}^2+\frac{2m}{r^2}\dot{t}^2+f^{-1/2}H. \label{tdd-bound1}\ee
On the right hand side, the terms $(1-\frac{m}{r})f^{-1/2}H$, $\frac{2m}{r^2}$ and $H$ all have finite positive limits as the geodesic approaches $r=2m$. As $\dot{t}$ and $f^{-1/2}$ both diverge to $+\infty$ in the limit, it follows that the first term on the right hand side (rhs) of (\ref{tdd-bound1}) dominates the other two, so that we may write
\be \ddot{t} < \hbox{[rhs of (\ref{tdd-bound1})]} \sim -(1-\frac{m}{r})f^{-1/2}\dot{t}^2,\quad s\to s_{2m}^+.\ee
Thus there exists $s_*\in(s_{2m},0)$ such that
\be \ddot{t}(s) < 0,\quad s<s_*.\ee
It follows by integrating twice that $s_{2m}>-\infty$: the geodesic reaches $r=2m$ in finite affine (or proper) time. \hfill$\square$


\section{Future evolution}

The main aim of this section is to build a comprehensive picture of the future evolution of radial null geodesics in expanding McVittie spacetimes with a big bang. We will also, where possible, draw conclusions about non-radial geodesics. The results we obtain both generalise and (attempt to) clarify previous results. The intention is to clarify the connection between the different global structures derived in previous work and the features of the corresponding background spacetimes (as encoded in the Hubble function $H$). The main results of the paper, which relate to bound photon and particle orbits, are presented in the following section and many of the results of this section are not required for those results. However this section is not wholly an aside: the reader interested in the results of Section VI should review Definitions 2 and 3, Lemmas \ref{lem:w(H)} and \ref{lemma-class2} and the paragraphs between Propositions \ref{prop:orng} and \ref{prop:irng-regular}.


As is evident from the contrasting results of \cite{nolan1999point} and \cite{kaloper2010mcvittie}, the asymptotic value as $t\to+\infty$ of the Hubble function $H(t)$ has a significant influence on the global structure of the spacetime. This is reflected in the results below, and to allow us to present those results in a clear manner, we define two subclasses of expanding McVittie spacetimes with a big bang. 

The first class comprises McVittie spacetimes where the scale factor of the isotropic background is that of an expanding, $k=0$, $\Lambda$-CDM isotropic universe. For the purposes of this paper, the defining properties are these:
\begin{definition} The spacetime with line element (\ref{eq:lel-mcv}) is a \textbf{Class 1 McVittie spacetime} if the Hubble function $H$ satisfies (i)-(iii) of Definition 1, plus the following conditions:
\begin{itemize}
\item[(iv)] $\lim_{t\to\infty}(H(t),H'(t),H''(t))=(H_0,0,0)$ where $H_0>0$.
\item[(v)] The background density and pressure satisfy (\ref{eq-rho0}) and (\ref{eq-press0}) with $\Lambda=3H_0^2$, and there is an equation of state $p_0=g(\rho_0)$ satisfying $g\in C^1[0,\infty)$ with the condition 
\be \kappa:=\frac32(1+g'(0))>0\label{sonic-cond2}\ee on the sound speed at zero density. 
\end{itemize}
\end{definition}

\begin{comment}
We note that the technical condition on the sound speed corresponds to the existence and positivity of the limit
$\lim_{\rho_0\to 0} 1 + \frac{p_0}{\rho_0}$ and so expresses a physically motivated energy condition, as well as a differentiability condition on $g$.
\end{comment}

The second class we consider corresponds to isotropic backgrounds that share features with spatially flat Robertson-Walker universes with equation of state $p=\hbox{constant}\times\rho$ and zero cosmological constant. The defining properties are these:

\begin{definition} The spacetime with line element (\ref{eq:lel-mcv}) is a \textbf{Class 2 McVittie spacetime}  if the Hubble function $H$ satisfies (i)-(iii) of Definition 1, plus the following conditions:
\begin{itemize}
\item[(iv)] $\lim_{t\to\infty}(H(t),H'(t),H''(t))=(0,0,0)$.
\item[(v)] The background density and pressure satisfy (\ref{eq-rho0}) and (\ref{eq-press0}) with $\Lambda=0$, and there is an equation of state $p_0=g(\rho_0)$ satisfying $g\in C^2[0,\infty)$ with the condition 
\be \kappa:=\frac32(1+g'(0))>0\label{sonic-cond3}\ee on the sound speed at zero density. 
\end{itemize}
\end{definition}

\begin{comment} The additional differentiability requirement on the equation of state function $\rho_0\mapsto g(\rho_0)$ is a technical condition required for some of the proofs of Section VI below.
\end{comment}

Our first result proves the existence of outgoing photon orbits that extend to infinity. This demonstrates the existence of an asymptotic region of every expanding McVittie spacetime with a big bang background where light rays extend to arbitrarily large radii.
\begin{proposition}\label{prop:orng}
Let $\gamma$ be an outgoing radial null geodesic of an expanding McVittie spacetime with a big bang background. Then $\gamma$ is future complete, and $\lim_{s\to\infty}(t,r)|_\gamma=(+\infty,+\infty)$.
\end{proposition}

\noindent\textbf{Proof:} Along an outgoing radial null geodesic, we have
\be \frac{dz}{dt}=\frac{H}{2}(1-z^2)+\frac{z}{4m}(1-z^2)^2,\ee
which is positive so that $z$ increases with $t$. Since $H>0$ (part (i) of Definition 1), we see that
\begin{eqnarray} \frac{dz}{dt}&>&\frac{z}{4m}(1-z^2)^2\nonumber\\
&>&\frac{z_0}{4m}(1+z_0)^2(1-z)^2 \quad \hbox{for all } t>t_0 \end{eqnarray}
where $z_0=z(t_0)$ and $t_0>0$ is arbitrary.  Integrating shows that $z\to 1$ (i.e.\ $r\to+\infty$) as $t\to+\infty$. From (\ref{ng1}) and part (iii) of Definition 1, we see that $\ddot{r}<0$ along the geodesic, and so the affine parameter $s$ must extend to $+\infty$ as $t,r$ do. \hfill$\square$

Next, we show that in Class 2 McVittie spacetimes, \textit{all} photon and particle orbits that originate outside a certain radius extend to infinity and are future complete. This and some subsequent proofs require some details of the horizon structure of McVittie spacetimes \cite{nolan1999point, kaloper2010mcvittie, lake2011more}. 

Recall that the \textit{horizon} of the McVittie spacetime with line element (\ref{eq:lel-mcv}) is the set of points with $\chi=g^{ab}\nabla_a r\nabla_b r=0$, the \textit{regular} region is the set of points with $\chi>0$ and the \textit{anti-trapped} region is the set of points with $\chi<0$. It is straightforward to show that the expanding condition $H>0$ leads to the region $\chi<0$ being anti-trapped rather than trapped. We will denote the regular region by $\Omega_R$, the anti-trapped region by $\Omega_A$ and the horizon by $\cal{H}$. We note that $\Omega$ is the disjoint union of $\Omega_R$, $\Omega_A$ and $\cal{H}$ and that 
\begin{eqnarray}
\Omega_R=\{(t,r)\in\Omega: H(t)<\frac{1}{r}(1-\frac{2m}{r})^{1/2}\},\label{reg-def}\\
\Omega_A=\{(t,r)\in\Omega: H(t)>\frac{1}{r}(1-\frac{2m}{r})^{1/2}\},\label{anti-def}\\
{\cal{H}}=\{(t,r)\in\Omega: H(t)=\frac{1}{r}(1-\frac{2m}{r})^{1/2}\}.\label{hor-def}
\end{eqnarray}

There are crucial differences between the horizons of Class 1 and Class 2 McVittie spacetimes. First, we note that in a Class 2 McVittie spacetime, all three sets $\Omega_A$, $\Omega_R$ and $\cal{H}$ are non-empty. However, in Class 1, this is not necessarily the case. Noting that 
\be H(t) > H_0 \quad\hbox{for all}\quad t>0\ee and that 
\be \frac{1}{r}(1-\frac{2m}{r})^{1/2}\leq \frac{1}{3\sqrt{3} m},\ee
we see that a necessary and sufficient condition for the existence of a horizon and a regular region in a Class 1 McVittie spacetime is that (\cite{kaloper2010mcvittie,lake2011more})
\be mH_0 < \frac{1}{3\sqrt{3}}.\label{hor-c1-exists}\ee
We note that this is identically the necessary and sufficient condition for the existence of horizons in the Schwarzschild-de Sitter spacetime with mass parameter $m$ and cosmological constant $\Lambda = 3H_0^2$. For Class 1 McVittie spacetimes, we restrict our attention to those cases where a horizon exists: 


\setcounter{definition}{1}
\begin{definition} \textrm{\textbf{[continued]}} 
\begin{itemize}
\item[(vi)] The horizon existence condition (\ref{hor-c1-exists}) is satisfied.
\end{itemize}
\end{definition}

From (\ref{hor-def}) and monotonicity of $t\mapsto H(t)$, we can describe the horizon by
\be t=t_h(r) = H_{\rm{inv}}(\frac{1}{r}\sqrt{1-\frac{2m}{r}}),\label{eq:hor-th}\ee
where $H_{\rm{inv}}$ is the inverse of the function $t\mapsto H(t)$: $H_{\rm{inv}}(H(t))=t$ for all $t>0$.
The function $r\mapsto t_h(r)$ has a positive global minimum at $r=3m$, is decreasing on $(2m,3m)$ and is increasing on $(3m,+\infty)$. We denote the global minimum by $t_{h,3m}$. Then we may also describe the two branches of the horizon by functions 
\be r_{h}^+:[t_{h,3m},+\infty)\to[3m,+\infty)\ee
and
\be r_{h}^-:[t_{h,3m},+\infty)\to(2m,3m]\ee
where by implicit differentiation we find 
\be \frac{d r_{h}^\pm}{dt} = \left.-r^2(1-\frac{3m}{r})^{-1}(1-\frac{2m}{r})^{1/2}H'(t)\right|_{r=r_h^\pm}. \label{horslope}\ee

In a Class 1 McVittie spacetime, $H(t)\to H_0>0$ as $t\to\infty$. Then on the horizon, $r$ is restricted to the interval $(r_-,r_+)$ where $r_\pm$ are the larger ($r_+$) and the smaller ($r_-$) of the two positive roots of $1-\frac{2m}{r}-r^2H_0^2=0$, and we have $t\to+\infty$ as $r\to r_\pm$ along the horizon. It is straightforward to prove that 
\be 2m<r_-<3m<r_+.\label{rpm-order}\ee From (\ref{horslope}), we see that on the \textit{inner branch}, $r_{h}^-$ decreases with $t$ from $r=3m$ to $r=r_-$, with $\lim_{t\to+\infty}r_{h}^-= r_-$. On the \textit{outer branch}, $r_{h}^+$ increases with $t$ from $r=3m$ to $r=r_+$, and $\lim_{t\to+\infty}r_{h}^+= r_+$. Note in particular that $\chi(t,r)<0$ for all $r>r_+$. See Figure 3.

In a Class 2 McVittie spacetime, we have $H_0=\lim_{t\to\infty}H(t)=0$. There is no restriction of $r$ - except of course that $r>2m$. From (\ref{horslope}), we see that on the \textit{inner branch}, $r_{h}^-$ decreases with $t$ from $r=3m$ to $r=2m$, and $\lim_{t\to+\infty}r_{h}^-= 2m$. On the \textit{outer branch}, $r_{h}^-$ increases with $t$ from $r=3m$ to $r=+\infty$, and $\lim_{t\to+\infty}r_{h}^+= +\infty$. See Figure 3.


\begin{figure}\label{horizons}
		{\includegraphics[scale=0.5]{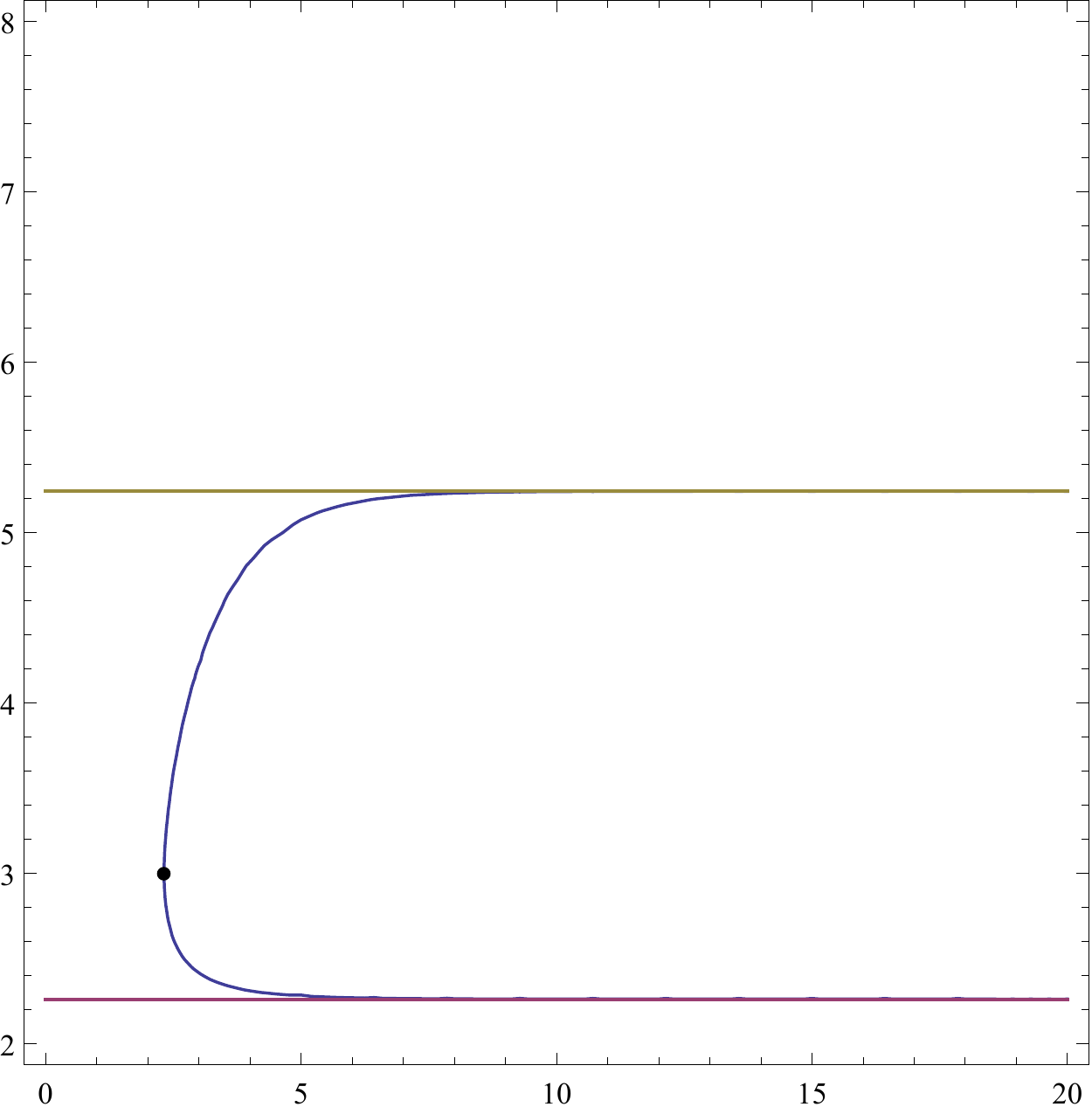} \vskip10pt \includegraphics[scale=0.5]{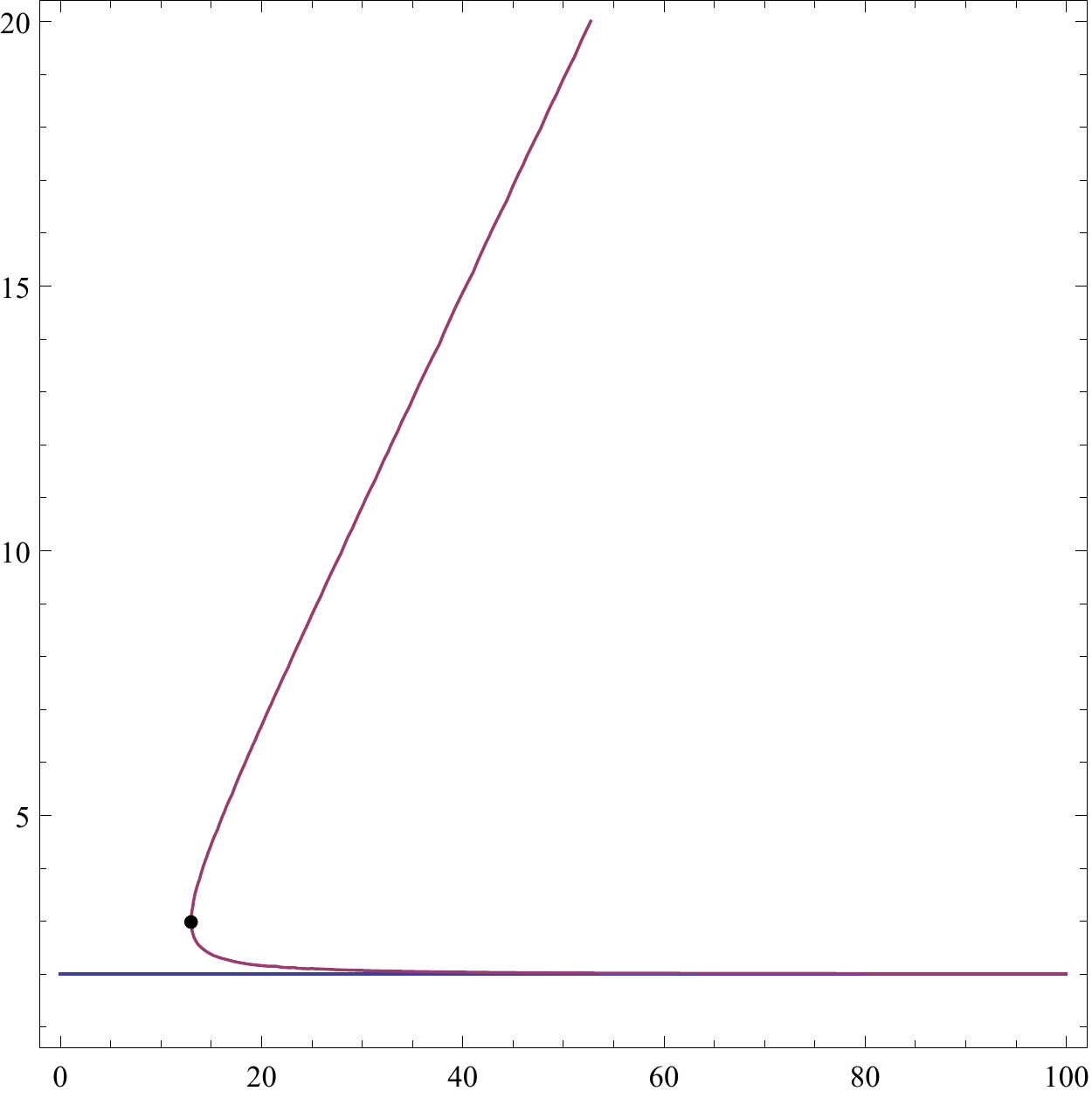}}
	\label{fig:CPO horizons}
	\caption{The horizon of (i) a Class 1 McVittie spacetime (top) and (ii) a Class 2 McVittie spacetime (bottom). In both cases, $t$ and $r$ are shown on the horizontal and vertical axes respectively, and $m=1$. The point at which the horizon first forms, $(t,r)=(t_{h,3m},3m)$, is also shown in both. For the Class 1 example, we have taken $H_0 = 0.15$ and $H(t)= H_0\coth(3H_0t)$. This puts the asymptotes of the horizon at $r=r_-=2.26$ (inner branch) and $r=r_+=5.24$ (outer branch). For the Class 2 example, we have taken $H(t)=2.5t^{-1}$. The asymptote of the inner branch of the horizon is $r=2m$, and $r\to\infty$ on the outer branch. In both cases, the regular region of spacetime ($\chi(t,r)>0$) is the convex region bounded above and below by the horizon. The exterior of this region is the anti-trapped region ($\chi(t,r)<0$).}
\end{figure}

The behaviour of ingoing radial null geodesics that are initially in the regular region of the spacetime is essentially the same in Class 1 and Class 2 McVittie spacetimes, as the next result shows. 


\begin{proposition}\label{prop:irng-regular} Let $\gamma$ be an IRNG of a Class 1 or a Class 2 McVittie spacetime with initial point $\gamma(0)$. If $\gamma(0)\in \Omega_R\cup{\cal{H}}=\{(t,r)\in\Omega:\chi(t,r)\geq0\}$, then there exists $s_0\in(0,\infty)$ such that $\lim_{s\to s_0^-} (t(s),r(s))=(+\infty,r_-)$ in Class 1, and $\lim_{s\to s_0^-} (t(s),r(s))=(+\infty,2m)$ in Class 2.
\end{proposition}

\noindent\textbf{Proof:} We use $r_{in}(t)$ to denote the solution of the IRNG equations considered as a curve in the $t-r$ plane and we have 
\be r_{in}'(t) = rf^{1/2}H(t)-f.\label{irng}\ee Essentially by definition, 
\be r_{in}'(t)|_{\cal{H}}=0.\ee
It follows from the description given above that the horizon acts as a one-way membrane with respect to IRNGs, injecting the trajectories into the regular region. Thus an IRNG with initial point in $\Omega_R$ remains in $\Omega_R$ on the maximal interval of existence. Furthermore, it follows that the geodesic cannot meet $\cal{H}$ again at some finite value of $t$ (this would violate uniqueness, as there is a unique IRNG at each point of the horizon which must enter rather than exit the regular region). It follows that $r_{in}'(t)<0$ on the maximal interval of existence and that $r>r_-$ along the geodesic. (We consider Class 1 explicitly. The proof is identical for Class 2, with $r_-$ replaced by $2m$.) Furthermore, equation (\ref{irng}) and the conditions on $H(t)$ show that this slope is bounded below, and so global existence in $t$ follows. By monotonicity, $\lim_{t\to\infty}r_{in}(t)\geq r_-$ exists, and it is clear that the limit must in fact equal $r_-$ (any other limit would yield a contradiction when $t\to\infty$ in (\ref{irng})). 

We prove that these geodesics have finite affine length to the future as follows. Using (\ref{ng2}) with the null condition $\epsilon=0$ and the ingoing condition $f\dot{r}=(rf^{-1/2}H-1)\dot{t}$ we obtain 
\be \ddot{t} = \left[-(1-\frac{m}{r})f^{-1/2}H+\frac{2m}{r^2}\right]\dot{t}^2\label{tdd-irng}\ee
along IRNGs. Given that the geodesic is (eventually) confined to $\Omega_R$, we have $\chi>0$ or equivalently $f^{-1/2}H>r^{-1}$ (see (\ref{reg-def})). This yields 
\be \ddot{t} > u(r)\dot{t}^2,\qquad u(r)=\frac{3m}{r^2}-\frac{1}{r}.\label{ddot-u}\ee
It is straightforward to show that $r\mapsto u(r)$ is a strictly decreasing function for $0<r<6m$: this condition holds along the geodesics in question for sufficiently late parameter times as $r\to r_-<3m$. Thus there exists $s_*<s_\omega$ such that (recalling $\dot{r}<0$ along these geodesics)
\be u(r(s)) > u(r(s_*))>0,\quad s\in(s_*,s_\omega).\ee
Then
\be \ddot{t} > u(r(s_*))\dot{t}^2,\quad s\in(s_*,s_\omega).\ee
Integrating twice shows that $t\to+\infty$ in finite affine time - that is, $s_\omega<+\infty$. \hfill$\square$

\begin{comment}
In \cite{nolan1999point}, which dealt with a subset of Class 2 McVittie spacetimes, it was erroneously claimed that the IRNGs of the previous result have infinite affine length to the future. A heuristic argument was put forth in \cite{kaloper2010mcvittie} correcting this claim, and generalising to Class 1 McVittie spacetimes. The result - finite affine length to the future - was proven rigorously in Appendix A of \cite{lake2011more}, using the argument repeated above. We have included the proof of finite affine length for purposes of continuity and clarity. 
\end{comment}

In Class 1, we can identify the following universal behaviour for all causal geodesics that are initially at sufficiently large radius. 


\begin{proposition}\label{prop:class1-causal-large}
Let $\gamma$ be a causal geodesic of a Class 1 McVittie spacetime with initial point $\gamma(0)$. If $r(0)=r|_{\gamma(0)}>r_+$, then $\gamma$ is future complete and $r\to\infty$ as $s\to\infty$. 
\end{proposition}

\noindent\textbf{Proof:} We note first that $\chi(t(0),r(0))<0$ where $t(0)=t|_{\gamma(0)}$. From (\ref{ng3}), we see that $\dot{r}(0)>0$, and so the conditions $r|_{\gamma(s)}>r_+$ and $\dot{r}(s)>0$ hold for $s\in[0,s_\omega)$, the right-maximal interval of existence for the geodesic. Hence $r(s)>r(0)>r_+>3m$ for all $s\in[0,s_\omega)$. It follows from (\ref{ng2}) that 
\be \ddot{t} < \left(1-\frac{2m}{r(0)}\right)^{-1/2}H(t(0)),\ee
where we have used $H'(t)<0$. Integrating shows that $\dot{t}(s)$ and $t(s)$ are bounded on finite intervals. From (\ref{ineq}) it follows that $\dot{r}(s)$ is bounded on finite intervals. Then a straightforward application of Theorem 1 (applied to the right-maximal rather than the left-maximal interval of existence) shows that $s_\omega=+\infty$, i.e.\ the geodesic $\gamma$ is future-complete. Since $\dot{r}(s)>0$ for all $s\in[0,\infty)$, the limit $r_\infty = \lim_{s\to\infty}r(s)\leq +\infty$ exists. If $r_\infty<+\infty$ (that is, if $r$ has a finite limit along the geodesic), then $\lim_{s\to\infty}\dot{r}=0$. From (\ref{ineq}), we obtain
\be -\chi\dot{t}<(\alpha+1)\dot{r}.\ee
Since $r(s)>r(0)>r_+$, we know that $\chi$ is negative and  bounded away from zero on the geodesic, and so $\dot{r}\to 0$ implies $\dot{t}\to 0$ as $s\to\infty$. But then taking the limit $s\to\infty$ in (\ref{ng3}) yields $\frac{\ell^2}{r_\infty^2}=\epsilon$. This yields a contradiction - and hence the conclusion that $\lim_{s\to\infty}r(s)=+\infty$ - unless $\ell=\epsilon=0$. So the theorem is proven except for the case of radial null geodesics. The case of outgoing RNGs is dealt with in Proposition \ref{prop:orng}. For ingoing RNGs satisfying the hypothesis of the theorem, we have the inequality
\be z(s) > z_+ = \sqrt{1-\frac{2m}{r_+}}.\ee
Then
\begin{eqnarray*}
\frac{dz}{dt} & > & \frac{(1-z^2)}{2}(H - \frac{z}{2m}(1-z_+^2)) \\
& = & \frac{(1-z^2)}{2}(H - \frac{z}{z_+}H_0)\\
&>& \frac{H_0}{2}(1-z^2)(1-\frac{z}{z_+}),
\end{eqnarray*}
using $H'(t)<0$ (so that $H(t)>H_0$) and $H_0=\frac{z_+}{2m}(1-z_+^2)$ which follows from the definition of $r_+$. Integrating (and using the initial condition $z(t_0)>z_+$) yields $z\to 1$ (i.e.\ $r\to+\infty$) as $t\to\infty$. \hfill$\square$

We can identify two further special cases of IRNGs in Class 1 McVittie spacetimes. 


\begin{proposition}\label{prop:class1-irng-special}
In a Class 1 McVittie spacetime,
\begin{itemize}
\item[(a)] there exists a 1-parameter  family of IRNGs with $(t,r)\to(+\infty,r_-)$ and $r(t)<r_-$ for all $t<+\infty$. These geodesics satisfy $\gamma(s)\in\Omega_A$ throughout their interval of existence and have finite affine length.
\item[(b)] There exists a unique IRNG with $(t,r)\to(+\infty,r_+)$; this geodesic is future-complete. 
\end{itemize}
\end{proposition}

To prove these results, we use a dynamical systems approach. It is useful towards this end to consider as independent variables $z(t)$ (i.e.\ the coordinate $z$ considered as a function of $t$ along the geodesic) and $\xi=H(t)$. The evolution of $z$ in terms of the state variables $(z,\xi)$ is given by (\ref{rngz}) (lower sign for IRNGs). The corresponding equation for the evolution of $H$ is furnished by the following lemma. This lemma is central to the analysis of Section VI, and so results for both Classes 1 and 2 are given here.  


\begin{lemma}\label{lem:w(H)}
In any expanding McVittie spacetime with a big bang background, there exists a function $w\in C^1([H_0,+\infty),\mathbb{R}_-)$ such that 
\be H'(t) = w(H(t)).\label{wdef}\ee In a Class 1 McVittie spacetime, this function satisfies
\be w(H_0)=0,\quad w'(H_0)=-2H_0\kappa, \label{w-props-c1}\ee
and in a Class 2 McVittie spacetime, 
\be w(0)=w'(0)=0,\quad w''(0)=-2\kappa,\label{w-props-c2}\ee
where $\kappa$ is the constant defined in part (v) of Definition 2 and Definition 3. Furthermore, this function extends to a function $w\in C^1(\mathbb{R},\mathbb{R})$.
\end{lemma}

\textbf{Proof:} Since $t\mapsto H(t)$ is monotone decreasing and $C^1$, this function has a monotone $C^1$ inverse $H_{\rm{inv}}$. Thus we can write
\be \xi = H(t) \Leftrightarrow t=H_{\rm{inv}}(\xi),\ee
where $H_{\rm{inv}}\in C^1((H_0,+\infty),\mathbb{R})$ and satisfies $\lim_{\xi\to H_0^+} H_{\rm{inv}}(\xi)=+\infty$ and $\lim_{\xi\to +\infty}H_{\rm{inv}}(\xi)=0$. Then we can write 
\be H'(t) = w(H(t))\ee
for the $C^1$ function $w=H'\circ H_{\rm{inv}}$. This proves existence of the required function $w$ and  shows that $w\in C^1((H_0,+\infty),\mathbb{R}_-)$, since we know by hypothesis that $H'(t)<0$ for all $t>0$. To prove the second part, we establish the equalities in limiting form (e.g.\ $\lim_{\xi\to H_0^+}w(\xi)=0$): this shows that $w$ extends to a function with the required properties. For convenience, we use the same name $w$ for this extension. 

To derive the required limits, we take the derivative of the background field equations (\ref{eq-rho0}) and (\ref{eq-press0}) with respect to $t$ (with $p_0=g(\rho_0))$ and eliminate 
\be \rho_0'(t)  = \frac{3}{4\pi}H(t)H'(t)\label{rho-prime}\ee to obtain
\be \frac{H''(t)}{H'(t)} = - 3H(t)(1+g'(\rho_0)). \label{hdd-over-hd}\ee
Taking the limit $t\to\infty$ and recalling that $\rho_0\to 0$ in this limit yields 
\be \lim_{t\to\infty} \frac{H''(t)}{H'(t)} = -2H_0\kappa, \ee
using Definition 2(v). On the other hand, we note that 
\begin{eqnarray} 
w'(\xi) &=& \frac{d}{dt}\left\{ H'(t) \right\} \frac{dt}{d\xi}\nonumber \\
&=& H''(t) \frac{dt}{dH} = \frac{H''(t)}{H'(t)}.
\label{w-prime}\end{eqnarray}
The result regarding the limit follows by noting that  $\xi\to H_0^+ \Leftrightarrow t\to+\infty$. Using Definition 2(iv), we have
\be \lim_{\xi\to H_0^+}w(\xi) = \lim_{t\to\infty}H'(t) = 0.\ee Thus the domain of $w$ may be extended to include $H_0$, and the required one-sided derivative exists to guarantee that $w\in C^1([H_0,\infty))$. 

This completes the proof for Class 1 McVittie spacetimes. For Class 2, the results $w(0)=w'(0)=0$ arise as a special case ($H_0=0$) of the Class 1 results, so it only remains to show that $w''(0)=-2\kappa$. 
Combining the $\xi$ derivative of (\ref{w-prime}) with the $t$ derivative of (\ref{hdd-over-hd}) and using (\ref{rho-prime}) yields
\be w''(\xi) = -3(1+g'(\rho_0))-\frac{9H^2}{4\pi}g''(\rho_0).\label{w-prime-prime}\ee
Taking the limit $\xi\to 0^+$ yields the required result, recalling the definition of $\kappa$ and that $g\in C^2[0,+\infty)$. 

To extend the domain of $w$ to the whole real line, we define $w(H_0-\xi)=-w(H_0+\xi)$ for $\xi>0$. \hfill$\square$ 

\vskip10pt
\noindent\textbf{Proof of Proposition \ref{prop:class1-irng-special}}
The equation governing IRNGs is (\ref{rngz}) with the lower sign. Using Lemma \ref{lem:w(H)} and taking $\xi=H(t)$, we can write this as a 2-dimensional dynamical system:
\begin{eqnarray}
\frac{dz}{dt} &=&\frac{\xi}{2}(1-z^2)-\frac{z}{4m}(1-z^2)^2,\label{ds-irng-z}\\
\frac{d\xi}{dt} &=& w(\xi).\label{ds-irng-xi}
\end{eqnarray}
Then (\ref{ds-irng-z}) and (\ref{ds-irng-xi}) define a $C^1$ dynamical system on $\mathbb{R}^2$. The points $P_1:(z,\xi)=(z_-,H_0)$ and $P_2:(z,\xi)=(z_+,H_0)$ are equilibrium points of this system, where $z_\pm:=z(r_\pm)$. Both are hyperbolic equilibrium points \cite{perkodifferential}, with eigenvalues 
\be \lambda_1 = -\frac{1}{4m}(1-z_\pm^2)(1-3z_\pm^2),\qquad \lambda_2 = -2H_0\kappa.\ee
Thus $\lambda_2<0$ in both cases, but $\lambda_1<0$ for $P_1$ and $\lambda_1>0$ for $P_2$. This follows from (\ref{rpm-order}) and the definition (\ref{eq:zdef}) of $z$. 

It follows that $P_1$ is an asymptotically stable equilibrium point: there exists a neighbourhood of $P_1$ in $\mathbb{R}^2$ such that every solution of (\ref{ds-irng-z}) and (\ref{ds-irng-xi}) with initial point in this neighbourhood satisfies $\lim_{t\to+\infty} (z(t),\xi(t)) = (z_-,H_0)$. Since the stable manifold is 2-dimensional, there are trajectories that approach the equilibrium point along every tangent at $P_1$. The 1-parameter family of the statement of the theorem correspond to trajectories of the dynamical system that approach the equilibrium point along directions with $z<z_-$. Thus these geodesics satisfy $z<z_-$ at late times, and so are confined to the anti-trapped region $\Omega_A$ at late times ($z_-<z$ for points in $\Omega_R\cup{\cal{H}}$). But $z'(t)>0$ for IRNGs in $\Omega_A$, so the geodesics are confined to $\Omega_A$ for all times. 

To complete the proof of part (a) of the proposition, it remains to prove that these geodesics have finite affine length. Proposition \ref{prop:causal-past} establishes that the geodesics have finite affine length to the past. To show finite affine length to the future, we note that the coefficient of $\dot{t}^2$ in the right hand side of (\ref{tdd-irng}) satsfies 
\be \lim_{s\to s_\omega} -(1-\frac{m}{r})f^{-1/2}H+\frac{2m}{r^2} = u(r_-),\ee
(see (\ref{ddot-u})) where $(s_\alpha,s_\omega)$ is the maximal interval of existence of the geodesic. Since $r_-<3m$, this coefficient is positive, and so there exists a positive constant $k^2$ and $s_*<s_\omega$ such that 
\be \ddot{t}>k^2 \dot{t}^2\quad\hbox{ for all } s\in(s_*,s_\omega).\ee
Integrating twice shows that $t$ diverges to $+\infty$ in finite $s$ completing the proof of part (a). 

For part (b), we note that $P_2$ is a saddle point of the dynamical system as $\lambda_1>0, \lambda_2<0$. The stable manifold theorem tells us that the nonlinear system has a 1-dimensional stable manifold $W^s$, tangent to the 1-dimensional stable space $E^s$ of the linearized system at $P_2$. This stable manifold corresponds to a unique IRNG satisfying $\lim_{t\to +\infty}(z,\xi)=(z_+,H_0)$. Along this geodesic, we have 
\be \lim_{s\to s_\omega} -(1-\frac{m}{r})f^{-1/2}H+\frac{2m}{r^2} = u(r_+)<0,\ee
since $r_+>3m$. Thus $\ddot{t}$ is eventually negative, and so cannot diverge to $+\infty$ in finite affine time. This geodesic is therefore future-complete. \hfill$\square$

\begin{comment}
In \cite{lake2011more}, the Penrose-Carter diagram for the McVittie spacetime with $H(t)= H_0\coth\left(\frac{3H_0 t}{2}\right)$ was constructed using a combination of analytical and numerical approaches. The results of Propositions \ref{prop:causal-past}-\ref{prop:class1-irng-special} show that the resulting structure applies for \textit{all} McVittie spacetimes of Class 1. This reinforces the results of \cite{kaloper2010mcvittie}, modulo the corrections of \cite{lake2011more}, e.g.\ the geodesic nature of the `cosmological horizon' $r=r_+(t)$ in the language and notation of \cite{kaloper2010mcvittie} (this is the geodesic $\eta_1$ of \cite{lake2011more} and is the unique IRNG of Proposition \ref{prop:class1-irng-special}(b) above) and the existence of the geodesics identified in Proposition \ref{prop:class1-irng-special}(a) above (also identified numerically in \cite{lake2011more}). We note also that the results establishing the affine lengths of the geodesics in Proposition \ref{prop:class1-irng-special} we first given in \cite{lake2011more}. 
\end{comment}

For Class 2 McVittie spacetimes, the asymptotic behaviour of ingoing radial null geodesics (IRNGs) is sensitive to the value of the constant $\kappa$ introduced in Definition 3. Proposition \ref{prop:irng-class2} generalises results of \cite{nolan1999point}: the following lemma is required for its proof. 

\begin{lemma}\label{lemma-class2}
In a Class 2 McVittie spacetime,
\be \lim_{t\to\infty} - \frac{H'(t)}{H(t)^2} = \kappa,\label{hp2lim}\ee
where $\kappa$ is the constant defined in part (v) of Definition 3.
\end{lemma}

\noindent\textbf{Proof:} From the Einstein equations (\ref{eq-rho0}), (\ref{eq-press0}) and Definitions 1 and 3, we see that $\lim_{t\to\infty}\rho_0=\lim_{t\to\infty}p_0=0$, and that 
\be \frac{p_0}{\rho_0}=-\frac23\frac{H'}{H^2}-1.\ee
Appealing to part (v) of Definition 3 and taking the limit $t\to\infty$ yields the result. \hfill$\square$


\begin{proposition}\label{prop:irng-class2} Let $\gamma$ be an ingoing radial null geodesic of a Class 2 McVittie spacetime, and let $\kappa$ be as defined in (\ref{sonic-cond3}).
\begin{itemize}
\item[(a)] If $\kappa<1$ and $\gamma(0)\in \Omega_A$ with $r(0)=r|_{\gamma(0)}$ sufficiently large, then $\gamma$ remains in $\Omega_A$ and is future complete with $\lim_{s\to\infty}(t(s),r(s))=(+\infty,+\infty)$;
\item[(b)] If $\kappa> 1$ and $\gamma(0)\in\Omega_A$, then $\gamma$ enters $\Omega_R$ and terminates at $\pom$ in finite parameter time. 
\end{itemize}
\end{proposition}

\noindent\textbf{Proof of part (a):}
We prove this result by identifying a family of curves that extends to infinity, and then show that each IRNG satisfying the hypothesis of the proposition must remain above a member of this family. 

We introduce a 1-parameter family $\{\Sigma_\nu:\nu>1\}$ of curves in $\Omega_A=\{(t,r)\in\Omega:\chi(t,r)<0\}$. These are defined to be the curves at each point of which the unique IRNG through that point has slope $r_{in}'(t)=\nu-1$. Then 
\be \Sigma_\nu = \{ (t,r)\in\Omega: H(t)=\frac{1}{r}(\nu-\frac{2m}{r})(1-\frac{2m}{r})^{-1/2}\}.\label{signudef}\ee
Notice that the horizon corresponds to $\Sigma_1$. It is straightforward to show that $\{\Sigma_\nu:\nu>1\}$ provides a foliation of $\Omega_A$, the anti-trapped region of the spacetime. Since $t\mapsto H(t)$ is monotone, we can invert (\ref{signudef}) to obtain a 1-parameter family of functions $t_\nu\in C^1((2m,+\infty),\mathbb{R}_+)$ such that 
\be (t,r)\in \Sigma_\nu \Leftrightarrow t=t_\nu(r).\ee
A straightforward calculation shows that $t_\nu'(r)>0$ for all $r>3m$ and all $\nu\geq 1$, and so there exists a 1-parameter family of functions $r_\nu\in C^1(\mathbb{R}_+,(3m,\infty))$ such that 
\be (t,r) \in \Sigma_\nu\cap\{(t,r)\in\Omega_A:r>3m\} \Leftrightarrow r=r_\nu(t).\ee
This amounts to saying that the set $\{(t,r)\in\Omega_A:r>3m\}$ may be foliation by $C^1$ curves in the $t-r$ plane, indexed by $\nu>1$, each one of which has the property that the (different) IRNGs intersecting that curve all have slope $r_{in}'(t)=\nu-1$, and we note that $r_\nu(t)\to+\infty$ as $t\to+\infty$ (on account of $H(t)\to0$ in this limit). 

We now show that $r_{in}(t)>r_\nu(t)$ for all $t>t_0$ and for an appropriate value of $\nu$. 

Implicit differentiation of the defining relation of $\Sigma_\nu$ in (\ref{signudef}) yields
\be \frac{dr_\nu}{dt}=-\left.\frac{H'}{H^2}\left(\nu-\frac{2m}{r}\right)^2\left(1-\frac{2m}{r}\right)^{1/2}\left(u_\nu(\frac{m}{r})\right)^{-1}\right|_{r=r_\nu(t)},\label{nuprime}\ee
where $u_\nu(x)=\nu-(\nu+4)x+6x^2$ and where we have repeatedly used (\ref{signudef}) to make convenient substitutions. It follows from Lemma \ref{lemma-class2} that 
\be \frac{dr_\nu}{dt}= \kappa\left( \nu + O(\frac{m}{r_\nu(t)})^2\right),\quad t\to\infty.\label{rnuslope}\ee
Then we can calculate 
\be \left.\frac{d}{dt}\left(\frac{r_{in}}{r_\nu}\right)\right|_{r_{in}=r_\nu}=\frac{1}{r_\nu(t)}\left(\nu-1-\kappa\nu + O(\frac{m}{r_\nu(t)})^2\right),\quad t\to\infty.\label{compare-slopes}\ee
As $\kappa<1$, we can choose $\nu>1$ such that $\nu-1-\kappa\nu>0$ (we require $\nu>1/(1-\kappa)$). From (\ref{compare-slopes}), we see that there exists $r_*>3m$ such that 
\be \left.\frac{d}{dt}\left(\frac{r_{in}}{r_\nu}\right)\right|_{r_{in}=r_\nu}>0\quad \hbox{for all } r_\nu(t)>r_*.\label{compare}\ee
Now consider the IRNG that passes through the point $(t_0,r_0)\in \Sigma_\nu$ with $r_0>r_*$. According to (\ref{compare}), the geodesic crosses $r=r_\nu(t)$ from below, and cannot cross this curve again for $t>t_0$ ($t\to r_\nu(t)$ is increasing). Thus $r_{in}(t)>r_\nu(t)$ for all $t>t_0$, and so $r_{in}(t)\to+\infty$ as $t\to\infty$ along the geodesic. By uniqueness, the same conclusion holds for all IRNGs satisfying $r|_{t_0}>r_0$. The usual convexity argument based on (\ref{ng1}) indicates that these geodesics have infinite affine length to the future. \hfill$\square$

\noindent\textbf{Proof of part (b):} Again, let $r_{in}(t)$ denote the IRNG in the $t-r$ plane, but with the additional condition that $r_{in}(t_0)=r_0$ where $(t_0,r_0)=(t,r)|_{\gamma(0)}\in\Omega_A$. Then we initially have $r_{in}^\prime(t)>0$. The IRNG thus either meets the horizon whereat $r_{in}'=0$ and then enters $\Omega_R$, or remains outside the horizon for all $t>t_0$. OUr aim now is to rule out the latter case. In this latter case, there exists $t_1>t_0$ such that $\lim_{t\to t_1}r_{in}(t)=\infty$ (this follows from the fact that $r\to +\infty$ on the relevant branch of the horizon). We may rule this out by the following convexity argument. First, we calculate that 
\be r_{in}''(t) = rf^{1/2}H'(t)+((1-\frac{m}{r})H(t)-\frac{2m}{r^2}f^{1/2})(rH(t)-f^{1/2}).\label{rinpp1}\ee
Since the geodesic remains in the anti-trapped region $\Omega_A$, we have $H>r^{-1}f^{1/2}$. This allows us to identify the dominant terms in (\ref{rinpp1}). If $t_1<+\infty$, we have
\be \frac{r_{in}''}{r_{in}}\sim H'(t_1)+H(t_1)^2,\qquad t\to t_1.\ee
But this rules out $r_{in}\to\infty$ in finite $t$. So we must have $t_1=+\infty$. In this case, we find
\begin{eqnarray*} \frac{r_{in}''}{r_{in}}&\sim& H'(t)+H(t)^2,\qquad t\to \infty \nonumber \\
&\sim & (1-\kappa)H(t)^2,\qquad t\to \infty
\end{eqnarray*} where we have used Lemma \ref{lemma-class2}. So by the hypothesis that $\kappa >1$, we see that $r_{in}''(t)<0$ for sufficiently large $t$. Now choose $t_2>t_0$ such that $r_{in}''(t)<0$ for all $t\geq t_2$, and let $\nu_2$ be the value of $\nu>1$ such that $(t_2,r_{in}(t_2))\in\Sigma_{\nu_2}$. We consider the set 
\be S=\{\nu\geq 1: r_{in}(t)>r_\nu(t) \hbox{ for all } t\geq t_2\}.\ee
This set is non-empty (it contains $\nu=\nu_2$) and is bounded below by $\nu=1$, since $\Sigma_1$ is the horizon which $r_{in}$ does not cross. Hence $S$ has a greatest lower bound $\nu_*$ with $1\leq \nu_* <\nu_2$. Then for any $\epsilon\in(0,\nu_2-\nu_*)$, $r_{in}$ crosses $\Sigma_{\nu_*+\epsilon}$ from below at some time $t_\epsilon$. At the point of crossing, $r_{in}'=\nu_*+\epsilon$, and by convexity, $r_{in}'(t)<\nu_*+\epsilon$ for all $t>t_\epsilon$. Now consider the surface $\Sigma_{\nu_*-\delta}$ (or $r=r_{\nu_*-\delta}$). From (\ref{rnuslope}), $r=r_{\nu_*-\delta}(t)$ has slope obeying 
\be r_{\nu_*-\delta}'(t) \sim \kappa(\nu_*-\delta),\quad t\to +\infty.\ee
Since $\kappa>1$, we can choose $\epsilon$ and $\delta$ so that $\kappa(\nu_*-\delta)>\nu_*+\epsilon$. This condition on the asymptotic slopes, along with the convexity property $r_{in}'(t)<\nu_*+\epsilon$ for all $t>t_\epsilon$, tells us that $r_{in}$ crosses $r_{\nu_*-\delta}$. This contradicts the infimum property of $\nu_*$, and so we rule out the possibility that $r_{in}$ does not cross the horizon. The evolution of the geodesic is thus subject to Proposition \ref{prop:irng-regular}. \hfill$\square$

\begin{comment} It is evident from the previous proposition that there is no general result governing the future evolution of causal geodesics in a Class 2 McVittie spacetime as we have in Proposition \ref{prop:class1-causal-large} for Class 1. At the very least, this behaviour depends on the value of $\kappa$. However, we can conclude that if $\kappa<1$, then a result corresponding to Proposition \ref{prop:class1-causal-large} applies. For $\kappa\geq 1$, we have not been able to draw any general conclusions, and it may be the case that no causal geodesics escape to infinity, other than the outgoing radial null geodesics covered by Proposition \ref{prop:orng}. 
\end{comment}

\begin{proposition}\label{prop:class2-causal} Let $\gamma$ be a causal geodesic of a Class 2 McVittie spacetime with $\kappa<1$. If $\gamma(0)\in \Omega_A$ and $r(0)=r|_{\gamma(0)}$ sufficiently large, then $\gamma$ remains in $\Omega_A$ and is future complete with $\lim_{s\to\infty}(t(s),r(s))=(+\infty,+\infty)$.
\end{proposition}

\noindent\textbf{Proof:} This is a straightforward application of the `radial confinement' result Lemma \ref{lem4}. As per Propositions \ref{prop:orng} and \ref{prop:class1-irng-special}, both the ingoing and outgoing radial null geodesics through $\gamma(0)$ remain in $\Omega_A$, are future complete and extend to infinite values of both $t$ and $r$. Thus $\chi$ is negative, and so by (\ref{ng3}) $\dot{r}$ is positive, along the maximally extended geodesic. From (\ref{ng2}), we then obtain 
\be \ddot{t} \leq (1-\frac{2m}{r(0)})^{-1/2}H(t(0)),\ee
where we have assumed (without loss of generality) that $r(0)>3m$, and we have used $-\epsilon\leq1$. Integrating shows that $\dot{t}$ remains finite for finite parameter times $s$, and then Lemma \ref{lem1} yields the same for $\dot{r}$. Applying Theorem 1 yields global existence for the geodesic, which, returning to the radial confinement result, completes the proof. \hfill$\square$


\section{Bound particle and photon orbits}

We now come to our main results: we prove that in any Class 1 or Class 2 McVittie spacetime, there are bound photon and particle orbits. That is, there are future-complete causal geodesics satisfying $r(s)<+\infty$ for all $s>0$. For photon orbits, we prove that these orbits approach $r=3m$ as $s\to\infty$. For particle orbits, we prove that the orbits are asymptotic to a radius corresponding to a stable circular orbit of the associated Scwarzschild(-de Sitter) spacetime (Schwarzschild-de Sitter for Class 1, Schwarzschild for Class 2). To prove the relevant results, we use a dynamical systems approach. The proofs run along somewhat different lines for the two classes, and so we consider them separately. However, the initial steps are the same for both classes. 

To begin, we write the geodesic equations as a dynamical system - that is, as a first order autonomous system of ODEs. We introduce the variables 
\be x=(x_1,x_2,x_3)=(H,r,\dot{r}),\quad y=\dot{t}. \label{c1vars} \ee
Then using (\ref{ng1}) and (\ref{ng2}) we can write the geodesic equations as a 4-dimensional system 
\be (\dot{x},\dot{y}) = G(x,y),\ee
for some function $G:\mathbb{R}^4\to\mathbb{R}^4$. 

The condition (\ref{ng3}) corresponds to a zero-order constraint:
\be (\ref{ng3}) \Leftrightarrow g(x,y)=0, \label{c1cona} \ee
where
\be g(x,y) := -\chi y^2 -2\alpha x_3 y + \lambda x_3^2 + \frac{\ell^2}{x_2^2}+|\epsilon| \label{gfndef}\ee
and we note that making the appropriate substitutions we have
\begin{eqnarray}
 \chi &=& f - x_1^2x_2^2,\label{chi-new}\\
\alpha &=& x_1x_2f^{-1/2}, \label{alpha-new}\\
\lambda &=& f^{-1}, \label{lambda-new} \\
f &=& 1-\frac{2m}{x_2}. \label{f-new}
\end{eqnarray}

The implicit function theorem allows us to solve $g(x,y)=0$ locally. Provided $\chi y +\alpha x_3\neq0$ at some point $P$, there is a neighbourhood of $P$ such that
\be g(x,y)=0 \Leftrightarrow y=u(x) \label{eq:y-u} \ee
for some $C^1$ function $u$. We use this to reduce the dimension of the problem. That is, we eliminate $y$ from $G(x,y)$ and write the geodesic equations as \be \dot{x} = F(x),\label{geo-ds-3dim}\ee
where $F:\mathbb{R}^3\to\mathbb{R}^3$ with components read off from (\ref{ng1}) and (\ref{ng2}), with the substitution $\dot{t}\to y=u(x)$.

One other technical manoeuvre is required in this process. We note that Lemma \ref{lem:w(H)} allows us to write $H'(t)=w(H(t))$ for the function $w$ introduced in that lemma, and so
\be \dot{x_1} = \frac{dH}{ds} = w(x_1)\dot{t}.\ee It is at this point that the methods of analysis diverge for the two classes. The difference arises due to the different structures of the function $w$ in the two cases. 

\subsection{Bound orbits in Class 1 McVittie spacetimes}

Throughout this subsection, we assume that our metric is that of a Class 1 McVittie spacetime as defined by Definitions 1 and 2. To begin, we recap some relevant results (including Lemma \ref{lem:w(H)}) that allow us to write the geodesic equations in dynamical systems form.
\begin{lemma}\label{lem7}
The geodesic equation for $x_1$ can be written as 
\be \dot{x_1} = w(x_1)u(x),\label{geo1}\ee
where $w$ and $u$ are as defined in (\ref{wdef}), (\ref{gfndef}) and (\ref{eq:y-u}). The function $w$  extends to a function $w\in C^1(\mathbb{R},\mathbb{R})$ satisfying
\be w(H_0) = 0,\quad w'(H_0) = -2H_0\kappa, \label{wpath0}\ee
where $\kappa$ is defined in part (v) of Definition 2, and we have
\be w(x_1)<0,\quad\hbox{ for all } x_1>H_0,\ee
\be w(H_0-\xi)  = - w(H_0+\xi),\quad\hbox{ for all } \xi>0.\ee  
\hfill{$\square$}
\end{lemma}


By making the appropriate substitutions in (\ref{ng2}), we can write down the other equations of the dynamical system:

\begin{eqnarray}
\dot{x_2} &=& x_3, \label{geo2}\\
\dot{x_3} &=& x_2f^{1/2}w(x_1)u(x)^2 + (1-\frac{3m}{x_2})\frac{\ell^2}{x_2^3}+\epsilon(\frac{m}{x_2^2}-x_1^2x_2).\nonumber\\
&&\label{geo3}
\end{eqnarray}

\subsubsection{Photon orbits in Class 1} 

We focus on the point $P:x=(H_0,3m,0)$. We find that 
\be \chi(P) =\chi_{|_{r=3m,H=H_0}} = \frac13-9m^2H_0^2 >0,\ee
and so $\chi$ is positive in a neighbourhood of $P$ (note the application here of the horizon existence condition (\ref{hor-c1-exists})). Recalling that $\dot{t}>0$ along all causal geodesics, we see that taking the appropriate solution of $g(x,y)=0$ in (\ref{c1cona}) yields
\be u(x) = u_1(x):= -\frac{\alpha}{\chi}x_3+\frac{1}{\chi}(x_3^2+\chi\frac{\ell^2}{x_2^2})^{1/2}.\label{u1sol}\ee
This is a smooth function of $x$ is a neighbourhood of $P$. For clarity, we collect some relevant results as follows. 
\begin{lemma}\label{lem8} In a neighbourhood of $P:x=(H_0,3m,0)$, the equations governing null geodesics of a Class 1 McVittie spacetime may be written as (\ref{geo2}), (\ref{geo3}) with $\epsilon=0$ and (\ref{geo1}) with $u=u_1$ and $u_1$ defined in (\ref{u1sol}). Furthermore, $P$ is an equilibrium point of this system. \hfill$|square$
\end{lemma}

Our main result is obtained by a relatively straightforward application of standard results on dynamical systems.
\begin{proposition}\label{prop:class1-bound-photon} In a Class 1 McVittie spacetime, there exists a 3-parameter family of null geodesics that are future complete, with $r\to 3m$ and $t\to+\infty$ as $s\to\infty$ along each member of the family. These geodesics correspond to bound photon orbits of the spacetime. 
\end{proposition}

\noindent\textbf{Proof:} Our starting point is the dynamical system 
\be \dot{x} = F(x) \label{ds-for-prop9}\ee
with 
\be F_1(x) = w(x_1)u_1(x),\qquad F_2(x) = x_3\ee
and
\be F_3(x) = x_2f^{1/2}w(x_1)u_1(x)^2 + (1-\frac{3m}{x_2})\frac{\ell^2}{x_2^3}.\ee
This is a $C^1$ dynamical system on a neighbourhood $\co_P\subseteq\mathbb{R}^3$ of $P$. We note that the projection 
\be \pi:\co_P\to \mathbb{R}^2:x=(x_1,x_2,x_3)\to (x_1,x_2)=(H,r)\ee
\textit{does not} project into $\Omega$, as we must include values of $H\in\mathbb{R}$ with $H<H_0$. Such points are not in the spacetime: the projection is into $\Omega$ if and only if $x_1=H>H_0$. 

The Jacobian matrix at $P$ is given by 
\be J(P) = \left( 
\begin{array}{ccc}
a & 0 & 0 \\
0 & 0 & 1 \\
b & c &0 
\end{array}
\right),
\ee
where
\begin{eqnarray} a &=& -\kappa^2H_0\frac{\ell}{m}(\frac13-9m^2H_0^2)^{-1/2},\\
b&=&-\sqrt{3}\kappa^2H_0\frac{\ell^2}{m}(\frac13-9m^2H_0^2)^{-1},\\
c&=&\frac{\ell^2}{(3m)^4}.
\end{eqnarray} 
This has eigenvalues $\sigma = a,\pm \sqrt{c}$, all of which are real, and so $P$ is a hyperbolic equilibrium point of the system. Two of the eigenvalues ($a,-\sqrt{c}$) are negative. Thus, by the Stable Manifold Theorem (see e.g.\cite{perkodifferential}), there exists a 2-dimensional differentiable manifold $W^s\subseteq\co_P$ such that any solution of (\ref{ds-for-prop9}) with $x(0)\in W^s$ satisfies $\lim_{s\to\infty} x(s) = P$. This manifold is tangent to $E^s$ at $P$, where $E^s$ is the 2-dimensional stable space of the linearized system
\be \dot{\delta x} = J(P)\delta x.\ee
By calculating the relevant eigenvectors, we find that 
\be E^s=\{x(P) + \xi \vec{v}_1 +\eta \vec{v}_2:\xi,\eta\in\mathbb{R}\},\label{Esform}\ee
where 
\be \vec{v}_1=((c+a^2)/b, 1, a),\qquad \vec{v}_2=(0,1,-\sqrt{c}).\ee

So far, these results apply only to the `unphysical' $\co_P$. To see the relevance to $\Omega$, and hence complete the proof, we note that those trajectories of $W^s$ which have tangent at $P$ corresponding to $\xi<0$ in (\ref{Esform}) satisfy $x_1(s)>H_0$ for all sufficiently large $s$ (note that $b<0$). These trajectories, of which there is a 3-parameter family (two parameters coming from the dimension of $W^s$ and $\ell$ providing the third) correspond to null geodesics of the spacetime which are future complete and have the asymptotic behaviour given in the statement of the proposition. 
\hfill$\square$

\begin{comment} We note that the orbits described in the previous result are asymptotic to the unique circular photon orbit $r=3m$ of the Schwarzschild-de Sitter spacetime with positive cosmological constant $\Lambda=3H_0^2$. They  are essentially as unstable as this unstable orbit at $r=3m$. The photon orbits of Schwarzschild-de Sitter spacetime are governed by the equation 
\be \ddot{r}=\left(1-\frac{3m}{r}\right)\frac{\ell^2}{r^3},\label{ng-sds}\ee
which is (\ref{ng1}) with the null condition $\epsilon=0$ and the Schwarzschild-de Sitter condition $H'(t)\equiv 0$. In terms of the dynamical system of Proposition \ref{prop:class1-bound-photon}, this equation corresponds to the invariant manifold $\{x:x_1=H_0\}$ of the flow. Then the circular photon orbit corresponds to the stationary point $(x_2,x_3)=(3m,0)$. This is a saddle of the 2-dimensional flow: thus, as is well know, the circular photon orbit of Schwarzschild-de Sitter spacetime is unstable. In terms of initial data, for a given $r(0)$ sufficiently close to $3m$, there are precisely (only) two values of $\dot{r}(0)$ for which the corresponding solution of (\ref{ng-sds}) lies on the stable manifold of the flow, giving $\lim_{s\to\infty}r(s)=3m$. All other values of $\dot{r}(0)$ yield trajectories which do not approach $r=3m$. In the non-static setting - that is, in the Class 1 McVittie spacetime - the stable manifold has an additional dimension. This corresponds to the flow of $H(t)$ to the asymptotic value $H_0$. However infinite fine tuning of the initial data is still required to ensure that the trajectory approaches $r=3m$: the stable manifold has co-dimension 1.
\end{comment}

\subsubsection{Particle orbits in Class 1}

For timelike geodesics, we find that there are bound orbits that are asymptotic to the stable circular particle orbits of the corresponding Schwarzschild-de Sitter spacetime. By `corresponding', we mean that the mass parameter $m$ and limiting Hubble value $H_0$ of the McVittie spacetime provide the mass parameter and positive cosmological constant $\Lambda=3H_0^2$ of the Schwarzschild-de Sitter spacetime. We recall that $r(s)=r_c$ is a circular orbit of Schwarschild-de Sitter spacetime if $G(r_c)=0$, where (see (\ref{ng1}) with $\epsilon=-1$ and $H(t)\equiv H_0$) 
\be G(r) = H_0^2r-\frac{m}{r^2}+(1-\frac{3m}{r})\frac{\ell^2}{r^3},\label{sds-circ}\ee
and that this orbit is stable if in addition
\be G'(r_c) < 0.\label{sds-circ-stable}\ee 
Stable circular orbits (or indeed circular orbits of any stability type) do not exist in Schwarzschild-de Sitter spacetime for all ranges of the parameters $m$ and $H_0$. We have the following result, which is proven in Appendix B.
\begin{proposition}\label{prop:isco} The Schwarzschild-de Sitter spacetime with mass parameter $m$ and positive cosmological constant $\Lambda=3H_0^2$ admits stable circular particle orbits if and only if $mH_0<\frac{2}{75\sqrt{3}}$. The radius of stable orbits satisfies $r_c\in(6m,r_+)$, where $r_+$ is the larger of the two positive roots of $1-2m/r-r^2H_0^2=0$.
\end{proposition}

\begin{comment} It is worth recalling what is meant by stability of these orbits. Let $m$, $H_0$ and $\ell$ be fixed. Write the timelike geodesic equations for Schwarzschild-de Sitter spacetime as the 2-dimensional dynamical system
\be \dot{r}=p,\qquad \dot{p}=(1-\frac{3m}{r})\frac{\ell^2}{r^3}-\frac{m}{r^2}+H_0^2r.\label{ds-sds}\ee
Let $r_c$ satisfy $G(r_c)=0$, so that $(r(s),p(s))=(r_c,0)$ is a circular orbit of the system. This orbit is said to be \textbf{stable} if for any $\epsilon>0$ there exists $\delta>0$ such that for all $(r_0,p_0)$ with $|(r_0,p_0)-(r_c,0)|<\delta$, the solution of (\ref{ds-sds}) with $(r(0),p(0))=(r_0,p_0)$ - that is, the unique geodesic through $(r_0,p_0)$ - satisfies $|(r(s),p(s))-(r_c,0)|<\epsilon$ for all $s\geq 0$. By using a Lyapunov function argument, it is straightforward to see that the orbit is stable if $G'(r_c)<0$.
\end{comment}

Corresponding to Lemma \ref{lem8}, we have this result:
\begin{lemma}\label{lem9}
Let $r_c>3m$ satisfy $G(r_c)=0$. In a neighbourhood of $Q:x=(H_0,r_c,0)$, the equations governing timelike geodesics of a Class 1 McVittie spacetime may be written as (\ref{geo2}), (\ref{geo3}) with $\epsilon = -1$ and (\ref{geo1}) with 
\be u=u_2:= -\frac{\alpha}{\chi}x_3+\frac{1}{\chi}(x_3^2+\chi(1+\frac{\ell^2}{x_2^2}))^{1/2}.\label{u2sol}\ee 
Furthermore, $Q$ is an equilibrium point and the hypersurface $x_1=H_0$ is an invariant manifold of this system. The restriction of the dynamical system to $x_1=H_0$ corresponds to the timelike geodesic equations of Schwarzschild-de Sitter spacetime.
\end{lemma} 

\noindent\textbf{Proof:} Collecting relevant definitions establishes the first part of the proof. For the second, we use Lemma \ref{lem:w(H)}, and we note that 
a straightforward calculation yields \be \left.\dot{x_3}\right|_{Q} = G(r_c).\ee We note also that
\be \chi_Q:=\left.\chi\right|_{r=r_c,H=H_0} = (1-\frac{3m}{r_c})(1+\frac{\ell^2}{r_c^2})>0.\label{chiq}\ee
\hfill$\square$

Corresponding to Proposition \ref{prop:class1-bound-photon}, we have this result which is established using centre manifold theory \cite{carr1981applications}. 
\begin{proposition}\label{prop:class1-bound-particle}
In a Class 1 McVittie spacetime with line element (\ref{eq:lel-mcv}), let $r_c$ satisfy $G(r_c)=0$ and $G'(r_c)<0$ where $G$ is defined in (\ref{sds-circ}). Then $Q:x=(H_0,r_c,0)$ is a stable equilibrium point of the timelike geodesic equations. Furthermore, there exists a neighbourhood $N_Q\subseteq \mathbb{R}^3$of $Q:x=(H_0,r_c,0)$ such that for any $x(0)\in N_Q$, there exists a positive constant $\gamma$ and a stable solution $\bar{x}(s)=(r(s),p(s))$ of (\ref{ds-sds}) such that 
\be x_1(s) = O(e^{-\gamma s}),\qquad (x_2(s),x_3(s))= \bar{x}(s)+O(e^{-\gamma s}),\quad s\to\infty\label{timelike-asymptotics}\ee 
where $(x_1(s),x_2(s),x_3(s))=x(s)$ is the unique timelike geodesic with initial data $x_0$.
\end{proposition}

\noindent\textbf{Proof:} A straightforward calculation of the Jacobian matrix corresponding to the equilibrium point $Q$ shows that it has one negative and two purely imagininary eigenvalues. Thus the equilibrium point has a one-dimensional stable manifold and a two-dimensional centre manifold. In fact the invariant manifold $x_1=H_0$ is a centre manifold of the point \cite{carr1981applications}. As noted in Lemma \ref{lem9}, the flow on the centre manifold is equivalent to the geodesic flow on Schwarzschild-de Sitter spacetime. Thus the equilibrium point is a stable point of the centre manifold. It follows by Theorem 2 of \cite{carr1981applications} and by the comments above that the equilibrium point is a stable point of the full flow (i.e.\ not restricted to the centre manifold), and that the asymptotic behaviour of the flow is as described in the statement of the proposition. 

\begin{comment} The interpretation of Proposition \ref{prop:class1-bound-particle} is as follows. The result tells us that for sufficiently late times ($H(t)$ close to $H_0$), timelike geodesics which are sufficiently close to $r=r_c$ with sufficiently small radial velocity ($\dot{r}$ close to 0), the orbit is equivalent to the corresponding orbit in Schwarzschild-de Sitter spacetime, modulo an exponentially decaying correction. These orbits are future complete, and correspond to bound particle orbits of the spacetime. Their past evolution is governed by Proposition \ref{prop:causal-past}. Note that to obtain geodesics of the spacetime, we must choose $x(0)\in N_Q$ with $x_1(0)>H_0$ (i.e. $H(t(0))>H_0$).
\end{comment}

\subsection{Bound orbits in Class 2 McVittie spacetimes}

We now consider Class 2 McVittie spacetimes as defined by Definitions 1 and 3. Again, we begin by writing down the result that gives a key geodesic equation in dynamical systems form. 
\begin{lemma}\label{lem10}
The geodesic equation for $x_1$ can be written as 
\be \dot{x_1} = w(x_1)u(x),\label{geo1cl2}\ee
where $w$ and $u$ are as defined in (\ref{wdef}), (\ref{gfndef}) and (\ref{eq:y-u}). The function $w$  extends to a function $w\in C^2(\mathbb{R},\mathbb{R})$ satisfying
\be (w(0),w'(0),w''(0)) = (0,0,-2\kappa), \label{w-lims-class2}\ee
where $\kappa$ is defined in part (v) of Definition 3, and we have
\be w(x_1)<0,\quad\hbox{ for all } x_1>0,\ee
\be w(-\xi)  = - w(\xi),\quad\hbox{ for all } \xi>0.\ee   \hfill$\square$
\end{lemma}


As in the Class 1 case, the other geodesic equations are given by (\ref{geo2}) and (\ref{geo3}).

\subsubsection{Photon orbits in Class 2}

The analysis above for photon orbits in Class 1 McVittie spacetimes follows through to the present case except in one important regard. The equilibrium point $P$ is now given by $x(P)=(0,3m,0)$, and the Jacobian matrix at $P$ is given by 
\be J(P) = \left( 
\begin{array}{ccc}
0 & 0 & 0 \\
0 & 0 & 1 \\
0 & c &0 
\end{array}
\right),\qquad c=\frac{\ell^2}{(3m)^4}.
\ee
Thus the eigenvalues are $0,\pm\sqrt{c}$, and the equilibrium point in non-hyperbolic \cite{perkodifferential}. (This follows from the fact that $w'(0)=0$, which is one of the features that distinguishes between Class 1 and Class 2.) Thus we must appeal to centre manifold theory \cite{carr1981applications}. For the linearized system at $P$, the centre space is given by $E^c=\{x:x_2=x_3=0\}$. By Theorem 1 of \cite{carr1981applications}, it follows that there is a local centre manifold $W^c$ of the nonlinear system which takes the form
\be x_2=h_2(x_1),\quad x_3=h_3(x_1),\quad h_i(x_1)=O(x_1^2),\quad x_1\to0, \label{centre-mfld}\ee
where $h_i$ is $C^2$ in a neighbourhood of $x_1=0$. The dynamics on the local centre manifold are governed by 
\be \dot{x_1}= w(x_1)u_1(x_1,h_2(x_1),h_3(x_1)).\ee
From Lemma \ref{lem10} and (\ref{centre-mfld}), we see that this reads
\be \dot{x_1} = -\kappa x_1^2 u_1(P) + O(x_1^3),\quad x_1\to 0.\label{cmfld-dyn}\ee
We note that (\ref{u1sol}) gives $u_1(P) = \frac{\ell}{\sqrt{3}m}>0$. It follows that if $x_1(0)$ is sufficently small and positive, and if $x(0)=(x_1(0), h_2(x_1(0)), h_3(x_1(0)))$ (that is, if the initial data for the geodesic correspond to a point of the centre manifold $W^c$), then $x_1(s)\to 0$ on the centre manifold as $s\to\infty$. 

In addition, there are local stable and unstable manifolds of the flow, $W^s$ and $W^u$ respectively. These are tangent at the equilibrium point to the stable and unstable spaces of the linearised system. The overall stability properties of the equilibrium point may be deduced from the topological equivalence of the nonlinear flow to the flow described by the partitioned system obtained by restricting to the local centre, stable and unstable manifolds respectively (see e.g. Theorem 2.7.2 of \cite{arrowsmith1990introduction}). Each of these is one dimensional, and given the description above of the stability properties of the flow on the local centre manifold, we conclude that we have the same result as applies in Class 1. Comment 8 following Proposition \ref{prop:class1-bound-photon} also holds in the present case. We note that in the language of \cite{kirchgraber1990geometry}, there is a two-dimensional local centre-stable manifold of the flow. This plays essentially the same role as the 2-dimensional stable manifold of Proposition \ref{prop:class1-bound-photon}. We summarise as follows. 
\begin{proposition}\label{prop:class2-bound-photon} In a Class 2 McVittie spacetime, there exists a 3-parameter family of null geodesics that are future complete, with $r\to 3m$ and $t\to+\infty$ as $s\to\infty$ along each member of the family. These geodesics correspond to bound photon orbits of the spacetime. \hfill$\square$
\end{proposition}

\subsubsection{Particle orbits in Class 2}

Turning to the case of timelike geodesics in Class 2 McVittie spacetimes, we encounter a technical difficulty which is not resolved by general dynamical systems theory: all of the eigenvalues of the Jacobian matrix at the equilibrium point have zero real parts. However, a useful feature of the dynamical system allows us to circumvent this problem and show that the stability properties of timelike geodesics in Class 1, as described in Proposition \ref{prop:class1-bound-particle}, carry over to Class 2. This feature is that we can think of the dynamical system as being a perturbation of the system describing the timelike geodesics of Schwarzschild spacetime. This allows us to exploit the existence of a conserved energy in this latter system. This energy (or more accurately, the effective potential, which plays the role of a Lyapunov function) is of course no longer conserved in the non-static McVittie spacetime, but (i) it remains positive definite and (ii) its evolution can be controlled by a Gronwall type argument. These properties allow us to conclude that the equilibrium point $x=(0,r_c,0)$ is stable. 

We begin by recalling the form of the relevant geodesic equations in a neighbourhood of $x=(0,r_c,0)$, where $r_c$ corresponds to a circular particle orbit of the Schwarzschild spacetime with mass parameter $m$.
\begin{lemma}\label{lem11}
Define
\be V(r) = -\frac{m}{r}+\frac{\ell^2}{2r^2}-\frac{m\ell^2}{r^3},\label{V-def}\ee
and let $r_c>3m$ satisfy $V'(r_c)=0$. Then in a neighbourhood of $Q:x=(0,r_c,0)$, the equations governing timelike geodesics of a Class 2 McVittie spacetime may be written as 
\begin{eqnarray}
\dot{x_1} &=& w(x_1)u_2(x),\label{c2-x1}\\
\dot{x_2} &=& x_3,\label{c2-x2}\\
\dot{x_3} &=& -V'(x_2) +x_2x_1^2+x_2f^{1/2}(x_2)w(x_1)u_2(x)^2,\label{c2-x3}
\end{eqnarray}
where $u_2$ is defined in (\ref{u2sol}) and $w$ satisfies the conditions of Lemma \ref{lem10}.
Furthermore, $Q$ is an equilibrium point and the hypersurface $x_1=0$ is an invariant manifold of this system. The restriction of the dynamical system to $x_1=0$ corresponds to the timelike geodesic equations of Schwarzschild spacetime.\hfill$\square$
\end{lemma} 

\begin{comment}
With $V'(r_c)=0$, we see that $r=r_c$ corresponds to a circular orbit of the Schwarzschild spacetime obtained by setting $H(t)\equiv 0$ in (\ref{eq:lel-mcv}). It is worth considering briefly how stability of such orbits may be deduced. We define the Lyapunov function 
\be W(x_2,x_3) = \frac12x_3^2+V(x_2)-V(r_c).\label{W-lyap}\ee
If $x_2=r_c$ is a local minimum of $V$ (which holds if $V'(r_c)=0, V''(r_c)>0$), then there exists $\epsilon>0$ such that 
\be |(x_2-r_c,x_3)|<\epsilon \Rightarrow \left\{ 
\begin{array}{c}
W(x_2,x_3)\geq0, \\
W(x_2,x_3) = 0 \Leftrightarrow (x_2,x_3)=(r_c,0).
\end{array} \right.
\label{W-props}
\ee
By a simple calculation,
\be \dot{W} = 0,\label{W-dot}\ee
and so by Lyapunov's stability theorem (see e.g.\ \cite{perkodifferential}, p.\ 130), $(x_2,x_3)=(r_c,0)$ is a stable equilibrium point: given $\epsilon>0$ and sufficiently small, there exists $\delta>0$ such that initial data with 
\be |(x_2(0),x_3(0))-(r_c,0)|<\delta\ee
yield solutions with
\be |(x_2(s),x_3(s))-(r_c,0)|<\epsilon \quad \hbox{ for all } s\geq0.\ee
By (\ref{W-dot}), the flow is confined to the level set 
\be W(x_2,x_3) = W(x_2(0),x_3(0)).\ee
We note that $W$ is essentially (the square of) the conserved energy associated with the timelike Killing vector of the exterior Schwarzschild spacetime. 
\end{comment}

Moving away from the invariant manifold $x_1=0$, the flow no longer has a conserved energy. However, we can use the same function $W$ to obtain useful control over the flow. The overall strategy is to introduce a mollified dynamical system that is equivalent to the system of Lemma \ref{lem11} in a neighbourhood of $Q$. We use the properties of a mollified form of $W$ to prove global existence of solutions and stability of $Q$ in this system. Stability in the full system then follows as the systems coincide in a neighbourhood of $Q$. 

We begin by noting that the geodesic equations of Lemma \ref{lem11} may be written as 
\begin{eqnarray}
\dot{x_1} &=& -k_1x_1^2A(x),\label{x1-Aform}\\
\dot{x_2} &=& x_3,\label{x2-Aform}\\
\dot{x_3} &=& - V'(x_2) +k_2x_1^2B(x),\label{x3-Aform}
\end{eqnarray}
where $k_1,k_2$ are positive constants, $A,B$ are $C^2$ in a neighbourhood of $Q$, and $A(Q)=B(Q)=1$.

We introduce a mollifier $\psi\in C^\infty_0(\mathbb{R}^3,\mathbb{R})$ with the property that 
\be \psi(x) = \left\{ \begin{array}{lcl} 
1 & \hbox{if} & |x-x(Q)|<1,\\
0 & \hbox{if} & |x-x(Q)|\geq 2,
\end{array} \right.
\label{molldef}
\ee
and we introduce mollified versions of the coefficients of the dynamical system:
\be \aep(x)=A(x\psi(\frac{x}{\epsilon})),\quad \bep(x)=B(x\psi(\frac{x}{\epsilon})).
\label{moll-coefft}
\ee
Then $\aep$ and $\bep$ coincide with $A$ and $B$ respectively for $|x-x(Q)|<\epsilon$, and we have $\aep(x)=\bep(x)=1$ for $|x-x(Q)|>2\epsilon$.   We also introduce a mollified form of the potential $V$. We introduce a smooth, bounded function $\vep\in C^\infty_0(\mathbb{R},\mathbb{R})$ with 
\be \vep(x_2) = \left\{ \begin{array}{ll} V(x_2), & |x_2-r_c|<\epsilon, \\ c, & |x_2-r_c|>2\epsilon, \end{array} \right. \label{vep-def}\ee
with $c>V(r_c)$. It follows that $\vep-V(r_c)$ is smooth, bounded and nonnegative for all $x_2$, coincides with $V(x_2)-V(r_c)$ for $|x_2-r_c|<\epsilon$ and vanishes if and only if $x_2=r_c$.
It follows that solutions of the system 
\begin{eqnarray}
\dot{x_1} &=& - k_1 x_1^2\aep(x),\label{x1-moll}\\
\dot{x_2} &=& x_3,\label{x2-moll}\\
\dot{x_3} &=& -\vep'(x_2) + k_2x_1^2\bep(x)\label{x3-moll}
\end{eqnarray}
coincide with solutions of the original system (\ref{x1-Aform})-(\ref{x3-Aform}) when those solutions satisfy $|x-x(Q)|<\epsilon$. 
\begin{lemma}\label{lem12}
Let $x(s)$ be a solution of (\ref{x1-moll})-(\ref{x3-moll}) with $|x(0)-x(Q)|<\epsilon$ and $x_1(0)>0$. Then there exists $\delta\in(0,1)$ such that 
\be 0< x_1(s) < \frac{x_1(0)}{1+k_1(1-\delta)x_1(0)s} \quad \hbox{for all } s> 0.  
\label{x1-bound}
\ee
\end{lemma}

\noindent\textbf{Proof:}
From the properties of $\aep$, we can choose $\epsilon>0$ so that there exists $\delta\in(0,1)$ such that 
\be 1-\delta < \aep(x) <1 + \delta \qquad \hbox{for all } x\in \mathbb{R}^3.\ee
Then (\ref{x1-moll}) yields 
\be \frac{d}{ds}\left(\frac{1}{x_1(s)}\right) = k_1\aep(x(s)) > k_1(1-\delta),\ee
and integrating yields the result. \hfill$\square$

\begin{corollary}
Let $x(s)$ be a solution of (\ref{x1-moll})-(\ref{x3-moll}) with $|x(0)-x(Q)|<\epsilon$ and $x_1(0)>0$. Then $x_1(s)<x_1(0)$  and $|sx_1(s)|<k_1(1-\delta)$ for all $s>0$. \hfill$|square$
\end{corollary}

Now we define 
\be \wep(x_2,x_3) = \frac12x_3^2+\vep(x_2)-V(r_c). \label{wep-def}\ee
Then using (\ref{x2-moll}) and (\ref{x3-moll}), we have
\be \frac{d\wep}{ds} = k_2\bep(x)x_1^2x_3,\label{wep-dot}\ee
which has the formal solution
\be \wep(s)=\wep(0) + \int_0^s k_2\bep(x(\sigma))x_1^2(\sigma)x_3(\sigma)d\sigma,
\ee
where for convenience we write $\wep(x_2(s),x_3(s))=\wep(s)$. 
Introducing the same bound for $\bep$ as we did for $\aep$ we can write
\be |\wep(s)|\leq |\wep(0)|+k_2(1+\delta)\int_0^s x_1^2(\sigma)|x_3(\sigma)|d\sigma.
\label{wep-bound1}\ee
Since $\vep(x_2)-V(r_c)\geq 0$ for all $x_2\in\mathbb{R}$ (see (\ref{vep-def})), we have
\be \frac12 x_3^2(s)\leq\wep(s)\leq Z(s), \ee
where $Z(s)$ is the non-negative function on the right hand side of (\ref{wep-bound1}). 
Then
\begin{eqnarray}
Z'(s) &=& k_2(1+\delta)x_1^2(s)|x_3(s)|\nonumber\\
&\leq&\sqrt{2}k_1(1+\delta)x_1^2(s)(Z(s))^{1/2}.
\end{eqnarray}
Integrating and using Lemma \ref{lem12} yields
\begin{eqnarray}
(Z(s))^{1/2} &\leq& |\wep(0)|^{1/2} +\frac{1}{\sqrt{2}}\frac{k_2(1+\delta)x_1^2(0)s}{1+k_1(1-\delta)x_1(0)s}\nonumber\\
&\leq & |\wep(0)|^{1/2} + c_1x_1(0).\label{z-bound}
\end{eqnarray}
where
\be c_1=\frac{k_1(1+\delta)}{\sqrt{2}k_2(1-\delta)}>0.\label{c-def}\ee
Since $\wep(s)$ is positive semi-definite, and using the fact that $r=r_c$ is a local minimum of the potential, this yields the following \textit{a priori} bounds:

\begin{lemma}\label{lem13}
Let $x(s)$ be a solution of (\ref{x1-moll})-(\ref{x3-moll}) with $|x(0)-x(Q)|<\epsilon$ and $x_1(0)>0$. Then there exist positive constants $c_1, c_2$ and $c_3$ such that
\begin{eqnarray}
|x_2(s)| & \leq & c_2(|\wep(0)|^{1/2} + c_1x_1(0)),\qquad s\geq 0,\\
|x_3(s)| & \leq & c_3(|\wep(0)|^{1/2} + c_1x_1(0)),\qquad s\geq 0.
\end{eqnarray}
\hfill$\square$
\end{lemma}

Thus we have shown that if we choose initial data with $|x(0)-x(Q)|<\epsilon$, then Lemmas \ref{lem12} and \ref{lem13} show that solutions of the mollified system satisfy $|x(s)-x(Q)|<O(\epsilon)$ for all $s>0$. In conjunction with Theorem 1 above, this is sufficient to prove global existence for solutions of the mollified system. Since solutions of the original system coincide with solutions of the mollified system for small $|x-x(Q)|$, this proves stability of the equilibrium point of the full system. We summarise as follows. 

\begin{proposition}\label{prop:class2-bound-particle}
In a Class 2 McVittie spacetime with line element (\ref{eq:lel-mcv}), let $r_c$ satisfy $V'(r_c)=0$ and $V''(r_c)>0$ where $V$ is defined in (\ref{V-def}). Let $x(Q)=(0,r_c,0)$. There exists $\epsilon>0$ such that, if $|x(0)-x(Q)|<\epsilon$, with $x_1(0)>0$, then there exists a global, unique solution of the timelike geodesic equations (\ref{c2-x1})-(\ref{c2-x3}) with initial data $x(0)$. $|x(s)-x(Q)|<O(\epsilon)$ for all $s>0$, and $\lim_{s\to+\infty} x_1(s)=0$. \hfill$\square$
\end{proposition}

\begin{comment}
The geodesics described by this proposition are bound particle orbits of the Class 2 McVittie spacetime in question. The result shows that if a particle orbit, at sufficiently late times, is, at any instant, sufficiently close in $(r-\dot{r})$ phase space to a stable circular orbit of the corresponding Schwarzschild spacetime, then the orbit remains close for all future times. 
\end{comment}

\section{Conclusions}

McVittie spacetimes provide us with a conceptually very simple model of the gravitational field of a point mass embedded in an otherwise isotropic, spatially flat universe. By conceptual simplicity, we mean that we can write down simple conditions on a spacetime that lead uniquely to the McVittie metric (\ref{eq:lel-mcv}). See for example \cite{nolan1998point} and \cite{nandra2012effect}. The interpretation of the resulting spacetime is somewhat more complex. However, a clear picture has been established by this stage. In the case of an expanding spacetime, the hypersurface $\pom=\{(t,r):t>0,r=2m\}$ forms the past boundary of the spacetime: all causal geodesics originate at this singular surface at a finite time in the past. Outgoing rays (outgoing radial null geodesics) escape to infinity, and there are families of ingoing rays (ingoing radial null geodesics) that meet either a horizon (Class 1) or a singularity (Class 2) at a finite time in their future. In addition, we have shown that large, physically relevant classes of McVittie spacetimes can `capture' both photons and massive particles, and keep them in orbit at finite radius for all time. This reinforces the interpretation of McVittie spacetimes as representing a central massive object in an otherwise isotropic universe. One can interpret all the results on test-particle (massive and massless) motion in McVittie spacetimes in terms of a balance between the attraction of the central mass and the expansion of the cosmological background. The results presented here show that there exists a region of the spacetime where the central mass always wins out, subject to initial conditions on the motion of the test particle - as in the vacuum black hole case (with or without a cosmological constant). 

As noted, these results also lay the foundations for the study of accretion disks in McVittie spacetimes. We have essentially established the existence of ISCOs in these spacetimes, and our results also make it clear that significant gravitational lensing occurs. We speculate that this may provide a useful setting for the calculation of lensing effects by galactic clusters or other over-densities in the universe. Likewise, it is clear that the central mass of a McVittie metric significantly disturbs and traps a proportion of a cloud of photons that propagates through a region of spacetime sufficiently close to that central mass. It would be of interest to determine the effect of this central mass on CMB calculations, e.g.\ through the recalculation of distance relations \cite{bolejko2011inhomogeneous}. We note in this context that while the McVittie spacetime in spatially homogenous in the sense that $\rho=\rho(t)$, effects of inhomogeneity would be manifest in the angular diameter distance $D_A$, which satisfies
\be \frac{d^2D_A}{ds^2} = -(|\sigma|^2 + 4\pi(\rho+p)(u_ak^a)^2)D_A,\label{add}\ee
where $u^a$ is the fluid 4-velocity, $k^a$ is tangent to a bundle of null geodesics (with shear $\sigma$) emanating from the source and the deriative is along $k^a$. In this and other ways, we believe that McVittie spacetimes could play an interesting role in inhomogeneous cosmological modelling, by providing a conceptually simple model of an overdensity in an otherwise spatially homogeneous universe. 


\begin{acknowledgments}
I thank Kayll Lake for interesting discussions of McVittie spacetimes. 
\end{acknowledgments}

\appendix

\section{Global structure of McVittie spacetimes admitting a circular photon orbit.}

For convenience, we recall the following facts. A McVittie spacetime has line element
\be ds^2=-(f-r^2H^2)dt^2-2rHf^{-1/2}dtdr+f^{-1}dr^2+r^2d\Omega^2,\label{app:lel-mcv}\ee
where $H=H(t)$ is the Hubble function of the background and $f=1-2m/r$. As proven in Section 3 above, the spacetime with this line element admits a circular photon orbit if and only if
\be H(t) = -H_0\tanh(AH_0t),\quad t\in \mathbb{R}.\label{app:hsol}\ee
where
\be A= \frac{\left(1-\frac{3m}{a}\right)}{\sqrt{1-\frac{2m}{a}}},\quad H_0=a^{-1}\sqrt{1-\frac{2m}{a}}\label{app:ah0def}
\ee
and the orbital radius is $r=a>2m$. The scale factor is given by 
\be S(t) = (\cosh(AH_0t))^{-1/A},\label{app:asol}\ee
which immediately shows that an important role is played by the sign of $A$. See Figures 1 and 2 in Section 3 above. We exclude the case $a=3m$ (which corresponds to Schwarzschild-de Sitter spacetime) and note that $A<0\Leftrightarrow a<3m$. 

Defining $z=\sqrt{1-2m/r}$ (cf.\ (\ref{eq:zdef})), the equations governing radial null geodesics are
\be \frac{dz}{dt}=\frac{H}{2}(1-z^2)\pm\frac{z}{4m}(1-z^2)^2.\label{app:rngz}\ee
Compactified representations of the outgoing (upper sign) and ingoing (lower sign) RNGs are shown in Figures 4 and 5. These diagrams also include the horizons of the spacetimes, given by 
\be H^2(t) = \frac{z^2}{4m^2}(1-z^2)^2.\label{hor-z}\ee

\subsection{CPO at radius $a<3m$.}

In Figure 4, we see that outgoing geodesics originate in the past at either (i) $(z,t)=(0,t_0)$ with $t_0>0$; (ii) $(z,t)=(z_-,-\infty)$; (iii) $(z,t)=(+\infty,-\infty)$ or (iv) $(z,t)=(z_+,-\infty)$. Note that $z_\pm$ are the larger and smaller of the two roots in $(0,1)$ of the asymptotic horizon equation
\be H_0^2 = \frac{z^2}{4m^2}(1-z^2)^2.\label{hor-asymp}\ee
There is a unique outgoing radial null geodesic (ORNG) in case (iv), which is a separatrix for cases (ii) and (iii). The existence of this geodesic may be established using dynamical systems, where $(z,t)=(z_+,-\infty)$ corresponds to a saddle. The ORNGs are either (v) future-complete, terminating at $(z,t)=(+\infty,+\infty)$ or (vi) of finite affine length, terminate at a finite time ($t<0$) in the future at $z=0$ ($r=2m$). The symmetry $H(-t)=-H(t)$ and the form of the RNG equations (\ref{app:rngz}) shows that the ingoing RNGs are simply the time-reversed outoing RNGs (as is manifest in Figure 4). By appealing to the geodesic equation (see (\ref{ng1}) with $\ell=\epsilon=0$)
\be \ddot{r} = rf^{1/2}H'(t)\dot{t}^2,\label{app:rdd}\ee
it is readily established that the boundary $r=2m$ is at finite affine distance in cases (i) and (v). Likewise, $z=z_-$ is at a finite affine distance to the past in case (ii), and geodesics extending to $r=+\infty$ (cases (iii) and (iv)) have infinite affine length in the relevant temporal direction. 

It is straightforward to establish the following facts in relation to the corresponding background spacetime - that is, the isotropic spacetime with line element (\ref{eq:lel-rw1}) with $H(t)$ given by (\ref{app:hsol}).
\begin{itemize}
\item The comoving world-lines $x=x_0$ are complete timelike geodesics, and $t$ is proper time along these geodesics. (Recall that $x=rS^{-1}(t)$ where $S$ is the scale factor.)
\item Outgoing RNGs are future complete and ingoing RNGs are past complete; $r$ and $|t|$ become infinite in the limit of infinite affine length. 
\item Outgoing RNGs either meet the regular origin at finite affine time in the past, or are past complete. 
\item Ingoing RNGs either meet the regulat origin at finite affine time in the future, or are future complete. 
\item The comoving coordinate $x$ has a finite limit as the affine parameter $|s|\to\infty$ along null geodesics. 
\end{itemize}

These points give rise to the conformal diagrams of Figure 5 for the McVittie spacetime and its background.

\begin{figure}[h]\label{cpo-rng-a-neg}
	\centering
		{\includegraphics[scale=0.25]{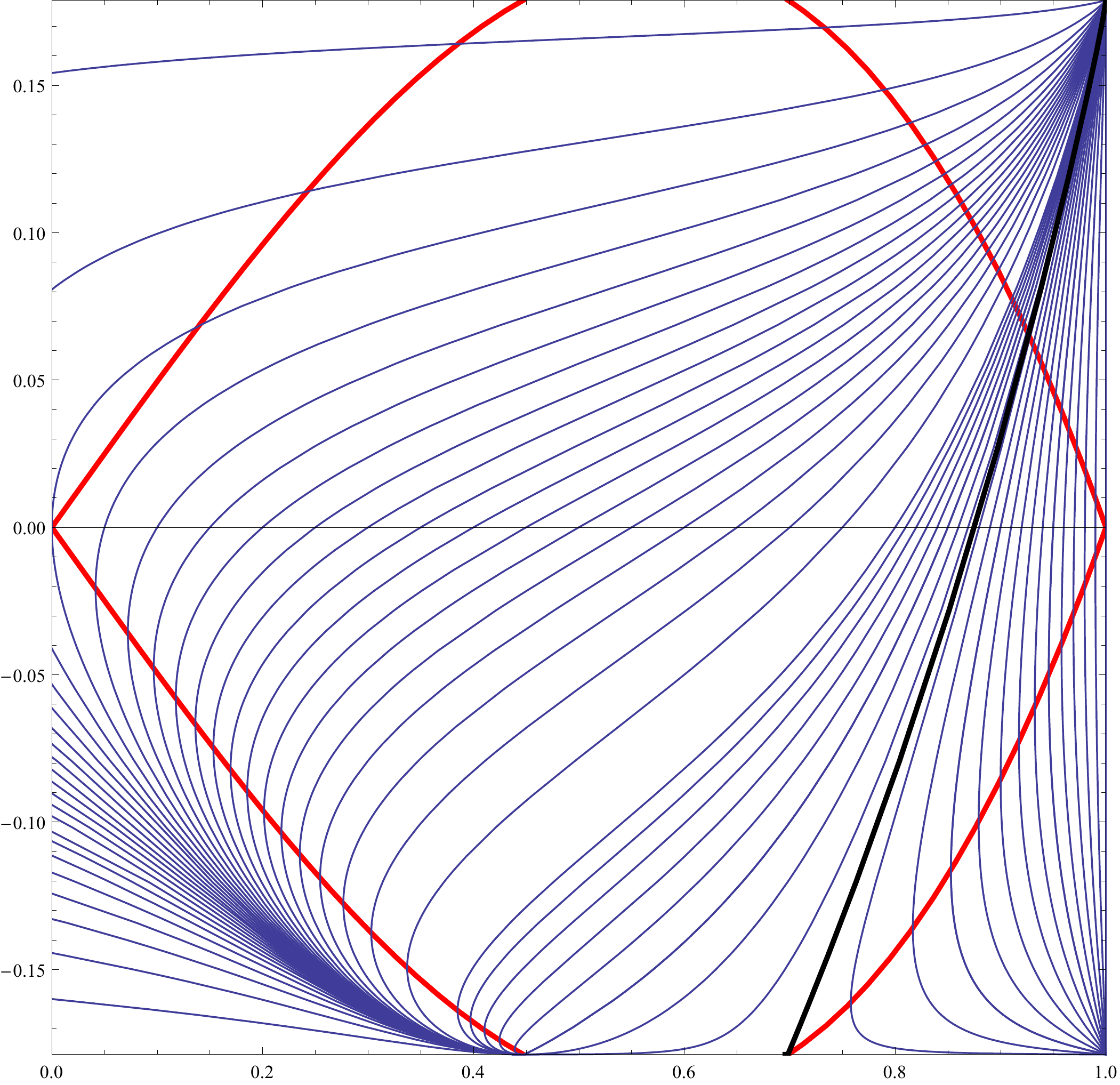}
		\vskip10pt \includegraphics[scale=0.25]{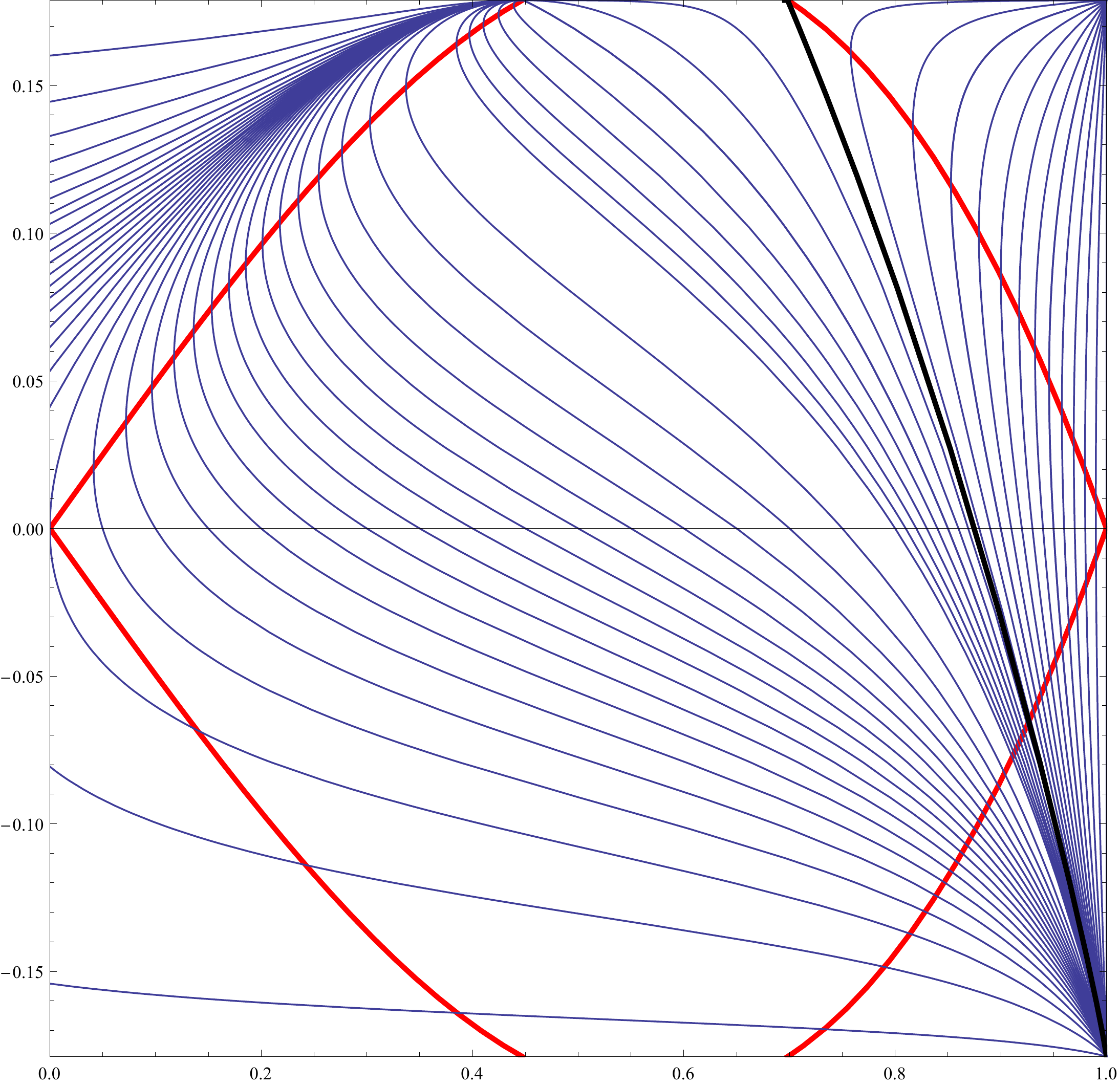}}
	\caption{Radial null geodesics and the horizon (red) of a McVittie spacetime admitting a circular photon orbit at radius $a<3m$. Here $m=1$ and $a=2.5$, giving $A=-0.447214<0$ and $H_0=0.178885$. Outgoing geodesics are shown in the upper diagram, IRNGs in the lower. Due to time reversal properties of the spacetime, IRNGs are simply time reversed ORNGs: the corresponding diagram is obtained by inversion. In the diagram, the radial coordinate is $z=\sqrt{1-2m/r}\in(0,1)$ and runs along the horizontal axis. Thus $r=2m$ at $z=0$ and $r\to\infty$ at $z=1$. The vertical axis corresponds to the value of $H(t)$ along the geodesics.Referring to Figure 1, we note that $t$ increases from bottom to top, giving the direction of flow along the geodesics. The critical geodesics $\bar{\gamma}_O$ and $\bar{\gamma}_I$ are shown in black: these are the unique ORNG/IRNG that originate/terminate at $r=r_+$. The trapping properties of the different regions of the spacetime may be read from this diagram: for $t<0$, there are trapped regions (both ORNGs and IRNGs move to the left - decreasing $r$ - with increasing $t$) and regular regions (ORNGs move right, IRNGs move left). For $t>0$, there are anti-trapped regions (both families move right). }
\end{figure}

\begin{figure}[h]\label{cpo-conf-a-neg}
\includegraphics[scale=0.7]{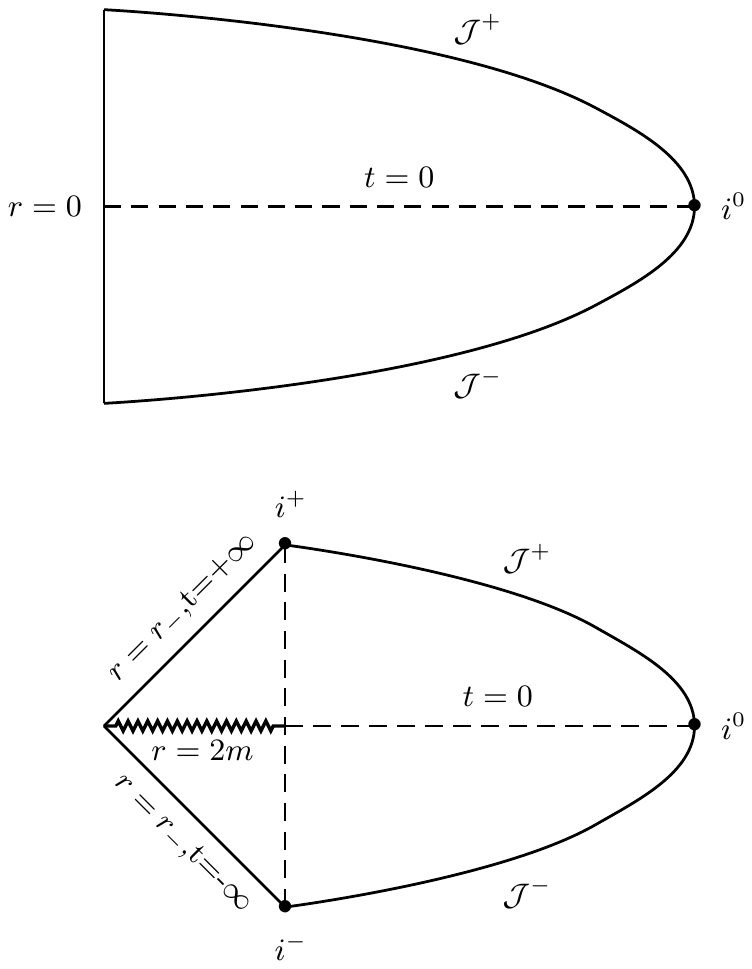}
\vskip -300pt
\caption{Conformal diagrams for the McVittie spacetime (bottom) that admits a circular photon orbit at $r=a<3m$, and the corresponding isotropic background (top). In both cases, $t$ increases in the vertical direction and ${\cal{J}}^\pm$ corresponds to $r=+\infty,t=\pm\infty$. The spacetime is collapsing ($H<0$) for $t<0$ and expanding ($H>0$) for $t>0$. In the McVittie case, we note the existence of ORNGs that (a) originate at $r=r_-$ and terminate at $r=2m$; (b) originate at $r=r_-$ and extend to $r=+\infty$; (c) originate at $r=+\infty, t=-\infty$ and extend to $r=+\infty, t=+\infty$ and (d) originate at $r=2m$ and extend to $r=+\infty, t=+\infty$. Similar results hold for IRNGs.} 
\end{figure}

\subsection{CPO at radius $a>3m$.}

As we see in Figure 6,  all ORNGs originate at $z=0$ ($r=2m$) at some time $t<0$. There is a unique ORNG which terminates at $(z,t)=(z_-,+\infty)$, and all others terminate at $z=0, t>0$ or at $(z,t)=(z_+,+\infty)$. The IRNGs are the time reversal of the ORNGs. Noting the exponential decay of $S(t)$ as $|t|\to\infty$ in this case (cf. (\ref{app:asol})), we find that the boundaries at $r=2m$ and $t=+\infty$ are at finite affine distance. 

\begin{figure}[h]\label{cpo-rng-a-pos}
	\centering
		{\includegraphics[scale=0.25]{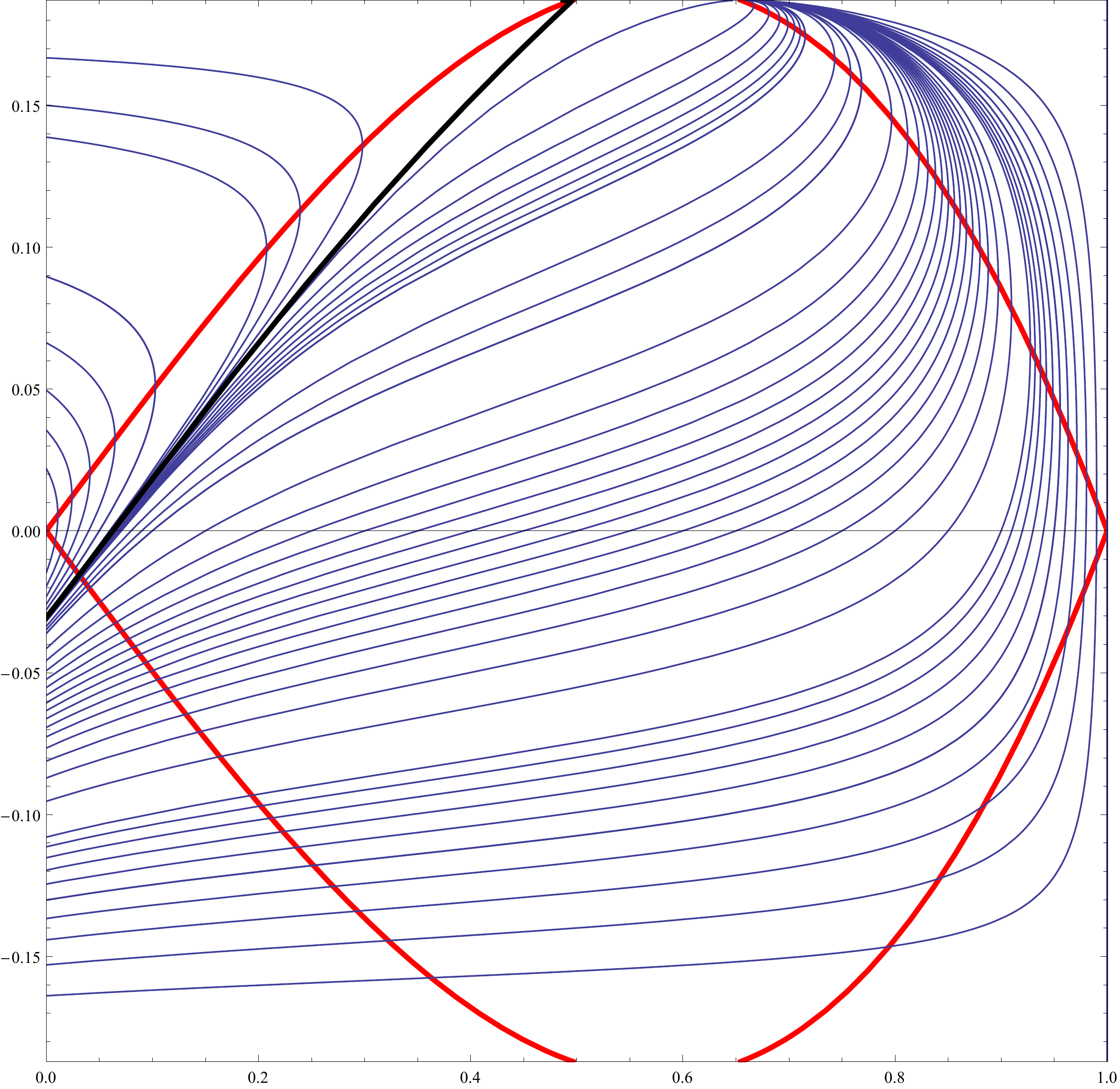} \vskip 10pt \includegraphics[scale=0.25]{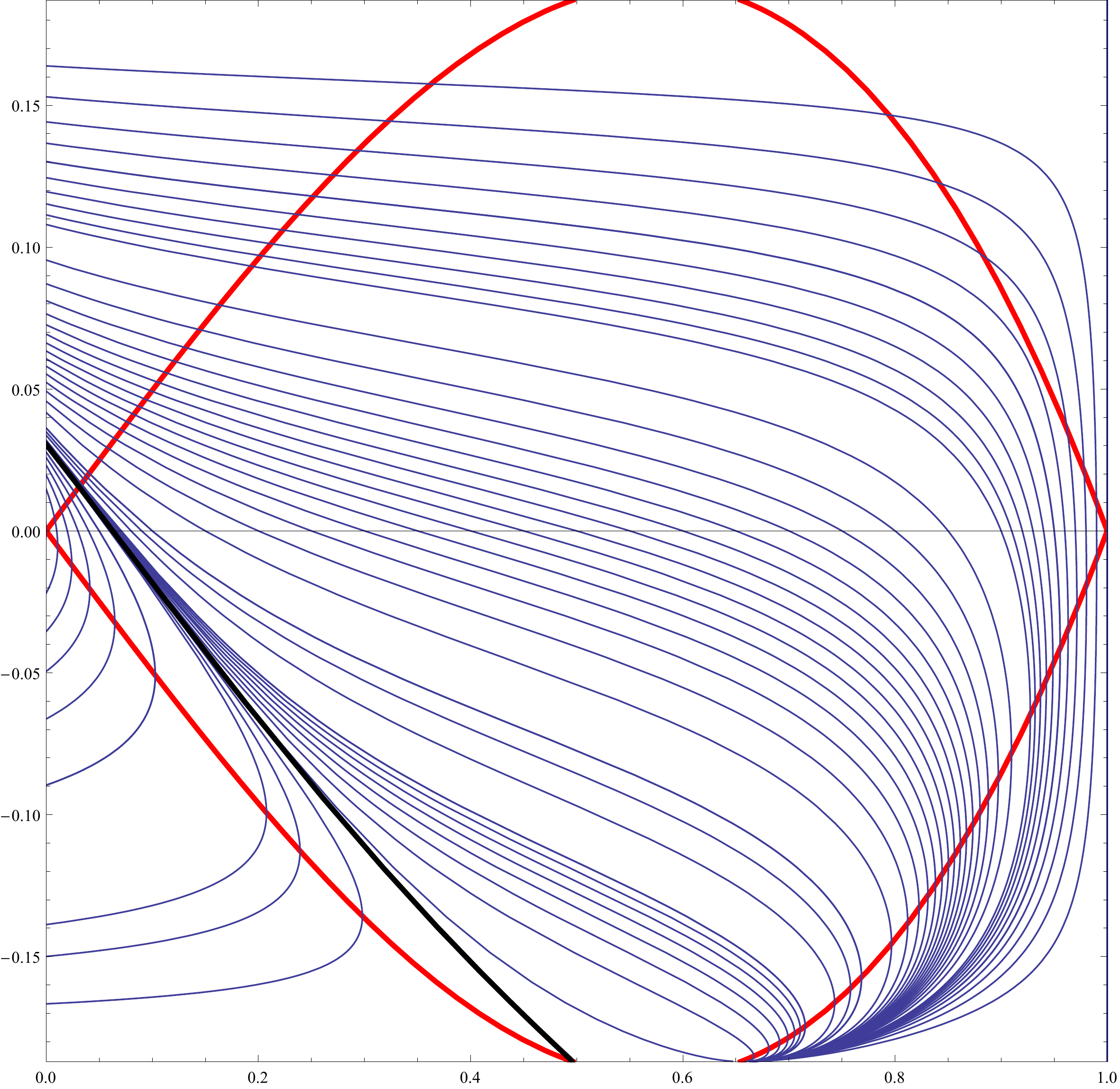}}
	\caption{Radial null geodesics of a McVittie spacetime admitting a circular photon orbit at radius $a>3m$. Here $m=1$ and $a=3.5$,giving $A=0.218218>0$ and $H_0=0.187044$. Outgoing geodesics are shown in the top panel, and ingoing in the bottom. In the diagram, the radial coordinate is $z=\sqrt{1-2m/r}\in(0,1)$ (see (\ref{eq:zdef}) below) and runs along the horizontal axis. Thus $r=2m$ at $z=0$ and $r\to\infty$ at $z=1$. The vertical axis corresponds to the value of $-H(t)$ along the geodesics. Referring to Figure 2, we note that $t$ increases from bottom to top, giving the direction of flow along the geodesics. The expanding region of the spacetime $(t<0)$ contains regular and anti-trapped regions, and the collapsing region $(t>0)$ contains regular and trapped regions.}
\end{figure}

The following properties hold in the background:
\begin{itemize}
\item The comoving worldlines are complete timelike geodesics carrying proper time $t$, and $r\to 0$ as $t\to\pm\infty$ along these geodesics. 
\item Along RNGs, $t$ becomes infinite in finite affine time, and $r$ remains finite in this limit.
\item All ORNGs meet $r=0$ at a finite $t$ and at finite affine distance in the future; all IRNGs meet $r=0$ in finite $t$ and finite affine distance in the past.  
\end{itemize}

The behaviour of the ORNGs as $t\to+\infty$ and the IRNGs as $t\to-\infty$ suggests that the spacetime is extendible across these coordinate boundaries. We note that the scale factor is asymptotic to the scale factor of a de Sitter spacetime:
\be S(t)= (\cosh(AH_0t))^{-1/A} \sim 2^{1/A}e^{-H_0 t},\qquad t\to+\infty,\ee
and so extending to de Sitter spacetime seems to be a natural possibility. We note that the time coordinate $t$ here shares features of the time coordinate $\hat{t}$ in the representation 
\be ds^2 = -d\hat{t}^2+\exp(2\alpha^{-1}\hat{t})(d\hat{x}^2+d\hat{y}^2+d\hat{z}^2)\ee
of the de Sitter metric. These coordinates cover only half of the full de Sitter hyperboloid. See Section 5.2 of \cite{hawking1973large}. We conjecture that a similar extension to Schwarzschild-de Sitter spacetime is possible in the McVittie case.

\begin{figure}[hh]\label{cpo-conf-a-pos}
\includegraphics[scale=0.7]{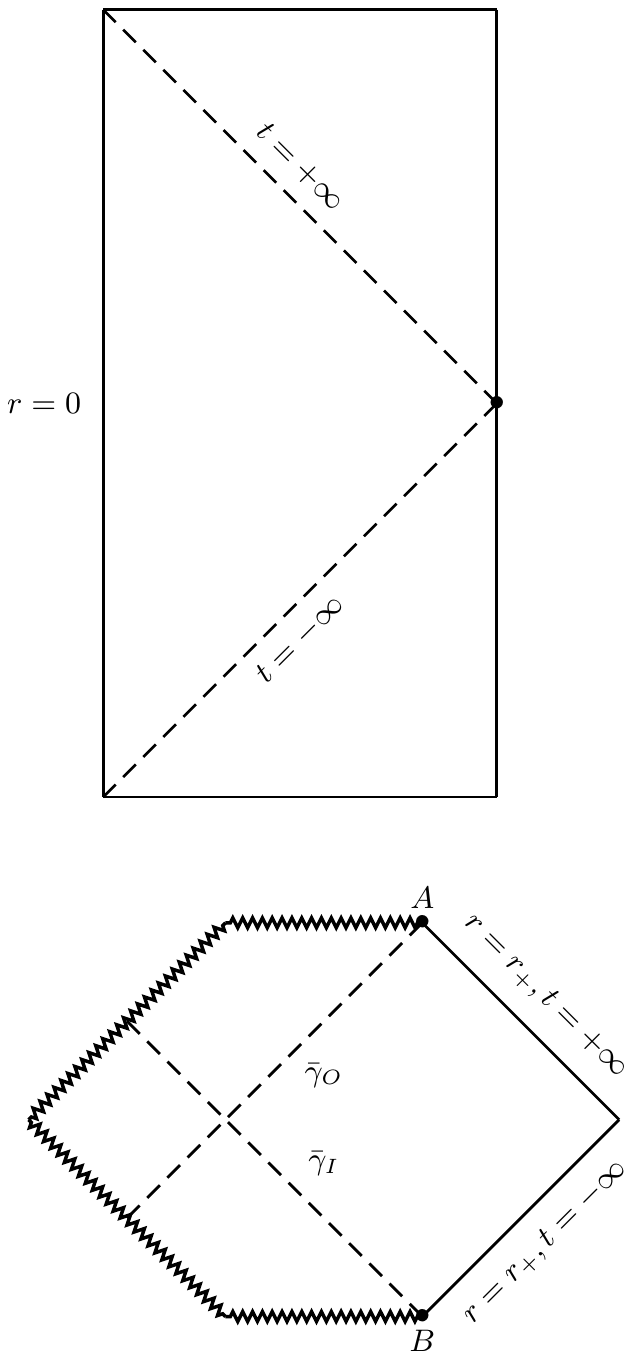}
\vskip -200pt
\caption{Conformal diagrams for the McVittie spacetime (bottom) that admits a circular photon orbit at $r=a>3m$, and the corresponding isotropic background (top). In both cases, $t$ increases in the vertical direction and ${\cal{J}}^\pm$ corresponds to $r=+\infty,t=\pm\infty$. The spacetime is expanding ($H<0$) for $t<0$ and collapsing ($H>0$) for $t>0$. In the McVittie case, the past singular boundaries (serrated) are at $r=2m, t<0$ and the future singular boundaries are at $r=2m, t>0$. At $A$, $(t,r)=(+\infty,r_-$ and at $B$, $(t,r)=(-\infty,r_-)$. We note the existence of (a) ORNGs that  originate at $r=2m$ and terminate at $r=2m$; (b) a unique ORNG $\bar{\gamma}_O$ that originates at $r=2m$ and extends to $r=r_+$ with $t\to+\infty$ and (c) ORNGs that originate at $r=2m$ and extend to $r=r_+$ with $t\to+\infty$. Similar results hold for IRNGs: $\bar{\gamma}_I$ is the unique IRNG that originates at $r=r_-$. We speculate that the spacetime extends to a Schwarzschild-de Sitter spacetime across $r=r_+, t=\pm\infty$, but we have not addressed this in the paper.} 
\end{figure}

The conformal diagrams that ensue from these features are shown in Figure 7. 

\section{Stable circular orbits in Schwarzschild-de Sitter spacetime.}
The conditions for a stable circular (particle) orbit in the Schwarzschild-de Sitter spacetime with mass parameter $m$ and cosmological constant $\Lambda=3H_0^2$ are $G(r_c)=0, G'(r_c)<0$ where $r_c$ is the orbital radius and $G$ is given in (\ref{sds-circ}). We wish to answer two related questions: for what range of the parameters $m,H_0$ do stable circular orbits exist? What is the radius of the innermost stable circular orbit (ISCO)? The algebra is simplified by introducing variables $u>2,v>0$ and a parameter $h_0>0$ defined by $r=mu, \ell^2=m^2v$ and $mH_0=h_0$. Note that $G(r_c)=0, G'(r_c)<0$ if and only if $\bar{G}(r_c)=0, \bar{G}'(r_c)<0$ where $\bar{G}(r)=r^4G(r)$ (recall that $r>2m$). This allows us to deal with polynomial functions only. Then the condition for a circular orbit is
\be \alpha(u,v):=h_0^2u^5-u^2+(u-3)v=0,\label{al-isco-def}\ee
and the condition for stability of this orbit is 
\be \beta(u,v):= 3u^2-4uv+15v <0.\label{be-isco-def}\ee
Note that we have used (\ref{al-isco-def}) to eliminate the term in $h_0^2$ that naturally arises in $\beta(u,v)$. 

Considering $\beta$ as a parametrised quadratic in $u$ shows that we must have $v>\frac{45}{4}$ and that (\ref{be-isco-def}) holds if and only if 
\be u_-(v) < u < u_+(v) \label{u-v-bounds}\ee
where 
\be u_\pm(v) = \frac{2v}{3}\left(1\pm\sqrt{1-\frac{45}{4v}}\right).\ee
We note that $u_-'(v)<0$ for $v>\frac{45}{4}$ and that $\lim_{v\to+\infty}u_-(v)=\frac{15}{4}$. This yields the (non-sharp) lower bound $u>\frac{15}{4}$ for the radius of a stable circular orbit. We also have the upper bound $u_-(v)\leq u_-(\frac{45}{4})=\frac{15}{2}$.

Suppose that for a given value of $h_0$, there exists a point $(u_1,v_1)$ such that $\alpha(u_1,v_1)=0$ and $\beta(u_1,v_1)<0$ - that is, there is a stable circular orbit at $u=u_1$ with angular momentum parameter $v_1$. Define \be g(u,v)= \frac{u^2-(u-3)v}{u^5}\ee
so that $g(u,v)=h_0^2$ if and only if $\alpha(u,v)=0$. Along the curve $g(u,v)=h_0^2$ we have
have $\ds{\frac{dv}{du}}>0$ for $\beta(u,v)<0$. It follows that the minimum of $u$ on $g(u,v)=h_0^2$ for which $\beta<0$ occurs along $u=u_-(v)$. Furthermore any value of $h_0$ for which there exists points $(u,v)$ with $g(u,v)=h_0^2$ gives rise to a curve with this property. 

Substituting then yields
\be h_0^2 = \frac{\ui-6}{\ui^3(4\ui-15)},\ee
where $\ui$ represents the ISCO radius for a given (scaled) Hubble parameter $h_0$. We now have the improved lower bound $u>6$. A straightforward calculation shows that $h_0$ and $\ui$ increase together. If there exists stable circular orbits for a given value of $h_0$, then there exists an ISCO for this value of $h_0$. This yields the maximum value of $h_0^2$ for which a stable circular orbit exists: this corresponds to the (global) maximum ISCO value, which is the maximum of $u$ on $u_-(v)$, i.e.\ $u=15/2$. Hence
\be h_0|_{\rm{max}}=h_0|_{\ui=\frac{15}{2}}=\frac{2}{75\sqrt{3}}.\ee

\newpage
\bibliographystyle{unsrt}
\bibliography{mybib}

\end{document}